\documentclass[reprint,showkeys,amsmath,amssymb,aps]{revtex4-2}

\usepackage{graphicx}
\usepackage{dcolumn}
\usepackage{bm}
\usepackage{nicefrac}
\usepackage[textsize=tiny]{todonotes}
\usepackage{hyperref}

\renewcommand{\eqref}[1]{Eq.~(\ref{#1})}
\newcommand{\figref}[2]{Fig.~\hyperref[fig:#1]{#1(#2)}}
\newcommand{\secref}[1]{Sec.~\ref{#1}}

\newcommand{\kIB}{k_{\textit{I}\rightarrow\textit{B}}}
\newcommand{\kBI}{k_{\textit{B}\rightarrow\textit{I}}}
\newcommand{\zB}{z_\textit{B}}
\newcommand{\zI}{z_\textit{I}}
\newcommand{\zBT}{\tilde{z}_\textit{B}}
\newcommand{\zIT}{\tilde{z}_\textit{I}}
\newcommand{\pB}{\tilde{\rho}_\textit{B}}
\newcommand{\pE}{\tilde{\rho}_\textit{E}}
\newcommand{\pI}{\tilde{\rho}_\textit{I}}
\newcommand{\Dfv}{\Delta f_\textit{v}}
\newcommand{\Dfl}{\Delta f_\textit{l}}
\newcommand{\rhoB}{\rho_\textit{B}}
\newcommand{\rhoI}{\rho_\textit{I}}

\newcommand{\eu}{e^{\beta u}}

\newcommand{\pLR}{p_\textit{LR}}
\newcommand{\pRL}{p_\textit{RL}}
\newcommand{\fL}{f_\textit{L}}
\newcommand{\fR}{f_\textit{R}}
\DeclareMathOperator{\erf}{erf}

\begin{document}

\title{Tuning nucleation kinetics via nonequilibrium chemical reactions}
\author{Yongick Cho}
\author{William M. Jacobs}
\email{wjacobs@princeton.edu}
\affiliation{Department of Chemistry, Princeton University, Princeton, New Jersey 08544, USA}

\date{\today}

\begin{abstract}
  Unlike fluids at thermal equilibrium, biomolecular mixtures in living systems can sustain nonequilibrium steady states, in which active processes modify the conformational states of the constituent molecules.
  Despite qualitative similarities between liquid--liquid phase separation in these systems, the extent to which the phase-separation kinetics differ remains unclear.
  Here we show that inhomogeneous chemical reactions can alter the nucleation kinetics of liquid--liquid phase separation in a manner that is consistent with classical nucleation theory, but can only be rationalized by introducing a nonequilibrium interfacial tension.
  We identify conditions under which nucleation can be accelerated without changing the energetics or supersaturation, thus breaking the correlation between fast nucleation and strong driving forces that is typical of phase separation and self-assembly at thermal equilibrium.
\end{abstract}

\maketitle

In living systems, phase separation can occur at a nonequilibrium steady state (NESS) as opposed to thermal equilibrium~\cite{berry2018physical,weber2019physics}.
For example, in active intracellular condensates, biomolecules may be degraded or post-translationally modified by enzymes that couple conformational changes to the conversion of a chemical fuel, such as ATP, to chemical waste~\cite{soding2020mechanisms}.
Although chemically driven fluids can undergo phase transitions resembling those of equilibrium systems, the phase behavior can be much richer when the enzymes that drive the reactions preferentially localize to one phase or when chemical fuel gradients couple to the local density of the phase-separating molecules~\cite{bartolucci2021controlling,kirschbaum2021controlling,zwicker2022intertwined}.
For example, phase separation taking place at a NESS can exhibit qualitatively different features compared to thermal equilibrium, including suppressed coarsening, monodisperse phase-separated droplet size distributions, and even spontaneous droplet division~\cite{zwicker2015suppression,zwicker2017growth,wurtz2018chemical,li2020non}.

Driven chemical reactions can also affect the kinetics of phase transitions, although the extent to which kinetic pathways at a NESS differ from those at equilibrium is not well understood.
Not only do driven chemical reactions provide additional control parameters beyond temperature and concentration with which to control a phase transition, but they might also alter the mechanism of phase separation.
This possibility contrasts with the behavior of equilibrium phase-separating fluids, in which strong thermodynamic driving forces are typically necessary to initiate homogeneous nucleation at equilibrium unless the system is near a critical point~\cite{oxtoby1992homogeneous,sear2007nucleation}.
The consequences of this correlation between thermodynamics and nucleation kinetics are well appreciated in the context of molecular self-assembly, especially in cases where strong driving forces are associated with kinetic trapping~\cite{whitelam2015statistical,perlmutter2015mechanisms,rogers2016using,jacobs2016self,hensley2022self}.
In principle, living systems must contend with similar trade-offs in order to harness phase separation for biological functionality~\cite{shin2017liquid,shimobayashi2021nucleation}.

Here we show that driven chemical reactions provide a mechanism to alter the nucleation pathway of a nonequilibrium phase-separating fluid.
To build intuition, we first describe simulations of phase coexistence and nucleation in a model of a fluid with driven chemical reactions, and we identify the conditions under which nucleation at a NESS cannot be described by an equilibrium theory.
Then, by introducing a general theoretical framework, we show that the difference between equilibrium and nonequilibrium nucleation kinetics arises from a nonequilibrium interfacial tension between the phases.
Our theoretical results establish how emergent interfacial properties can tune the kinetics of phase separation and self-assembly far from equilibrium.

\begin{figure*}
  \includegraphics[width=\textwidth]{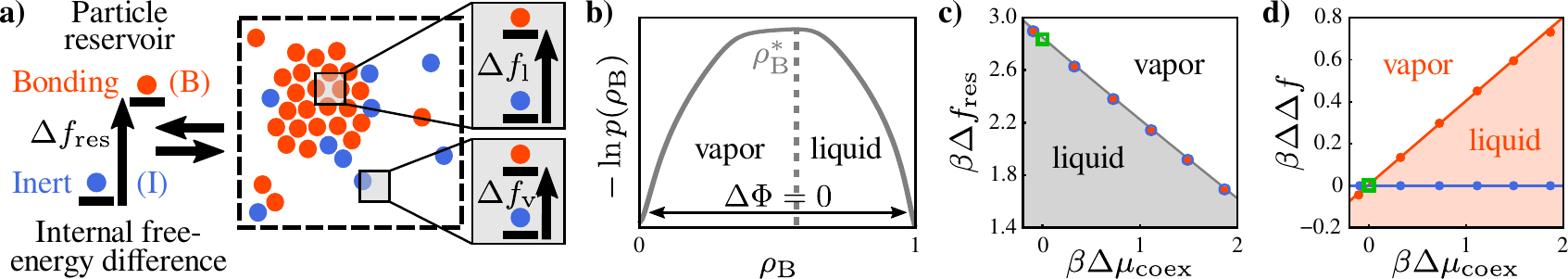}
  \caption{
  Simulating driven chemical reactions at a phase-separated NESS.
  (a)~Schematic of an open system with inhomogeneous chemical reactions.  The effective internal free-energy differences between the \textit{B} and \textit{I} states in the liquid and vapor phases are $\Dfl$ and $\Dfv$, respectively.
  (b)~An example steady-state distribution in an inhomogeneous model.
  (c)~Phase diagram for equilibrium (green), nonequilibrium homogeneous (blue), and inhomogeneous (orange) models, and
  (d)~quantification of the inhomogeneous reactions assuming $\beta\epsilon = -2.95$, $k^\circ = 10^{-1}$, and $\rho_\textit{v} = 0.05$ at coexistence. The shaded (unshaded) region indicates where liquid (vapor) is stable for all models in (c) and for the inhomogeneous model only in (d).}
  \label{fig:1}
\end{figure*}

In order to study nonequilibrium phase separation via molecular simulation, we adopt the framework of stochastic thermodynamics~\cite{seifert2012stochastic,van2015ensemble} and consider an open system connected to a particle reservoir [\figref{1}{a}].
For simplicity, we perform simulations using a two-dimensional square lattice model, in which empty lattice sites represent solvent.
We assume that the particles have two internal states: a bonding state (\textit{B}) that promotes phase separation due to attractive nearest neighbor interactions with bond energy $\epsilon < 0$, and an inert state (\textit{I}) that is isoenergetic to an empty lattice site.
The internal free-energy difference between the internal states in the reservoir is $\Delta f_{\text{res}}$, so that the fugacities of the two states are related by $\zB / \zI = \exp(-\beta \Delta f_{\text{res}})$, where $\beta \equiv (k_{\text{B}}T)^{-1}$.
Our model is closely related to the equilibrium lattice gas, which exhibits a first-order phase transition between a dilute vapor (\textit{v}) phase and a condensed liquid (\textit{l}) phase below a critical temperature~\cite{pathria1996statistical}.
However, unlike the equilibrium lattice gas, particle transitions between the system and the reservoir in our model may not obey time-reversal symmetry.
The product of rates for inserting a bonding particle, changing its internal state, and returning it to the reservoir may therefore differ from that of the reversed sequence by a factor $\exp(\beta\Delta\mu)$, where $\Delta\mu$ is the chemical potential difference used to drive reactions between the internal states inside the system (Appendix~\hyperref[app:model]{A}).
We assume that $\Delta\mu$ is uniform throughout the system.

Interconversion between \textit{B} and \textit{I} states can occur either directly or via exchange with the reservoir.
The ratio of the direct forward and backward $\textit{B} \rightleftharpoons \textit{I}$ reaction rates is controlled by $\Delta\mu$ in accordance with ``local detailed balance''~\cite{seifert2012stochastic,van2015ensemble}.
Meanwhile, the reservoir-mediated pathway is governed by $\Delta f_{\text{res}}$.
The steady-state populations are therefore influenced by the relative fluxes through these competing pathways, which can be tuned by specifying the rate for $\textit{I}\rightarrow\textit{B}$ transitions, $\kIB$.
If $\kIB$ is constant, then the chemical reactions are \textit{homogeneous}.
By contrast, if $\kIB$ is influenced by the local environment, then we refer to the reactions as \textit{inhomogeneous}.
In our lattice model, fluids with inhomogeneous chemical reactions have a $\kIB$ rate that depends on the nearest-neighbor particles and thus on the local potential energy.

To quantify inhomogeneous chemical reactions at a NESS, we introduce an \textit{effective} internal free-energy difference, ${\beta\Delta f \equiv -\ln(\rhoB / \rhoI) + \ln\langle\exp(-\beta\Delta u_{\textit{I}\rightarrow\textit{B}})\rangle_{\textit{I}}}$, where $\rhoB$ and $\rhoI$ are the steady-state number densities of particles in the \textit{B} and \textit{I} states, respectively, and the second term represents an average of the potential energy change due to converting an \textit{I} to a \textit{B} particle at steady state (Appendix~\hyperref[app:model]{A}).
At equilibrium, $\Delta f = \Delta f_{\text{res}}$.
At a NESS, an explicit dependence of $\kIB$ on the local potential energy causes $\Delta f$ to differ between the liquid and the vapor phases, such that $\Delta\Delta f \equiv \Dfl - \Dfv \ne 0$.
Although this mapping between nonequilibrium and equilibrium models is not exact in general, measuring $\Delta\Delta f$ provides crucial insight into the differences between fluids with inhomogeneous and homogeneous reactions.

We illustrate the differences between homogeneous and inhomogeneous chemical reactions by performing kinetic Monte Carlo simulations~\cite{gillespie2007stochastic} of a particular nonequilibrium fluid model.
We consider a fluid in which $\Delta f_{\text{res}} > 0$, meaning that the \textit{I} state is more populous in the vapor phase, while bonding stabilizes the \textit{B} state in the liquid phase [\figref{1}{a}].
We implement chemical reactions by assuming Markovian transitions and local detailed balance, such that reactions taking place inside the system are controlled by $\Delta\mu$ (Appendix~\hyperref[app:model]{A}).
We obtain homogeneous reactions if we set $\kIB$ equal to a constant, $k^\circ$, which represents the ratio between the timescales for internal state changes and particle diffusion.
To obtain inhomogeneous reactions, we assume that the $\textit{I}\rightarrow\textit{B}$ transition rate is a decreasing function of the local potential energy, $u$, at a lattice site (Appendix~\hyperref[app:model]{A}).
We emphasize that due to local detailed balance, this choice of $\kIB$ implies that \textit{both} the $\textit{I}\rightarrow\textit{B}$ and $\textit{B}\rightarrow\textit{I}$ rates are enhanced at low potential energy when positive chemical drive is applied, resulting in an increased $\Delta f$ in the liquid relative to the vapor phase [\figref{1}{a}].

We identify the conditions for nonequilibrium phase coexistence with both homogeneous and inhomogeneous reactions by equating the total probability of being in the vapor versus the liquid phase at steady state [\figref{1}{b}].
This is analogous to the equal pressure construction in equilibrium grand-canonical phase-coexistence simulations~\cite{wilding1995critical}, and implies that the open system transitions between the liquid and vapor phases with equal forward and backward rates.
To this end, we use a form of nonequilibrium umbrella sampling~\cite{warmflash2007umbrella} to calculate the steady-state probability as a function of the number density of bonding particles, $p(\rhoB)$~\footnote{\label{footnote:SI}See Supplementary Material for a detailed description, which includes \cite{noya2008} and \cite{hansen2013simpleliquids}}.
As is characteristic of a first-order phase transition, we observe a barrier with respect to $-\ln p(\rhoB)$ that scales with the lattice length $L$ as the system size is increased~\cite{chandler1987introduction}.
Based on the value of the order parameter $\rhoB^*$ at the top of this barrier, we determine the steady-state probabilities of the vapor and liquid phases, $p_{\textit{v}} \equiv \int_0^{\rhoB^*} p(\rhoB)d\rhoB$ and $p_{\textit{l}} \equiv \int_{\rhoB^*}^1 p(\rhoB)d\rhoB$, respectively.
We then define the dimensionless thermodynamic driving force between bulk phases to be $\beta\Delta\Phi \equiv L^{-2} \ln \left( p_\textit{l} / p_\textit{v} \right)$ and associate phase coexistence with $\Delta\Phi = 0$ [\figref{1}{c}].
As anticipated, measuring $\Delta\Delta f$ between coexisting phases confirms that only the potential-energy dependent choice for $\kIB$ results in inhomogeneous reactions, regardless of $\Delta\mu_\text{coex}$, the nonequilibrium drive at coexistence [\figref{1}{d}].

\begin{figure}
  \includegraphics[width=\columnwidth]{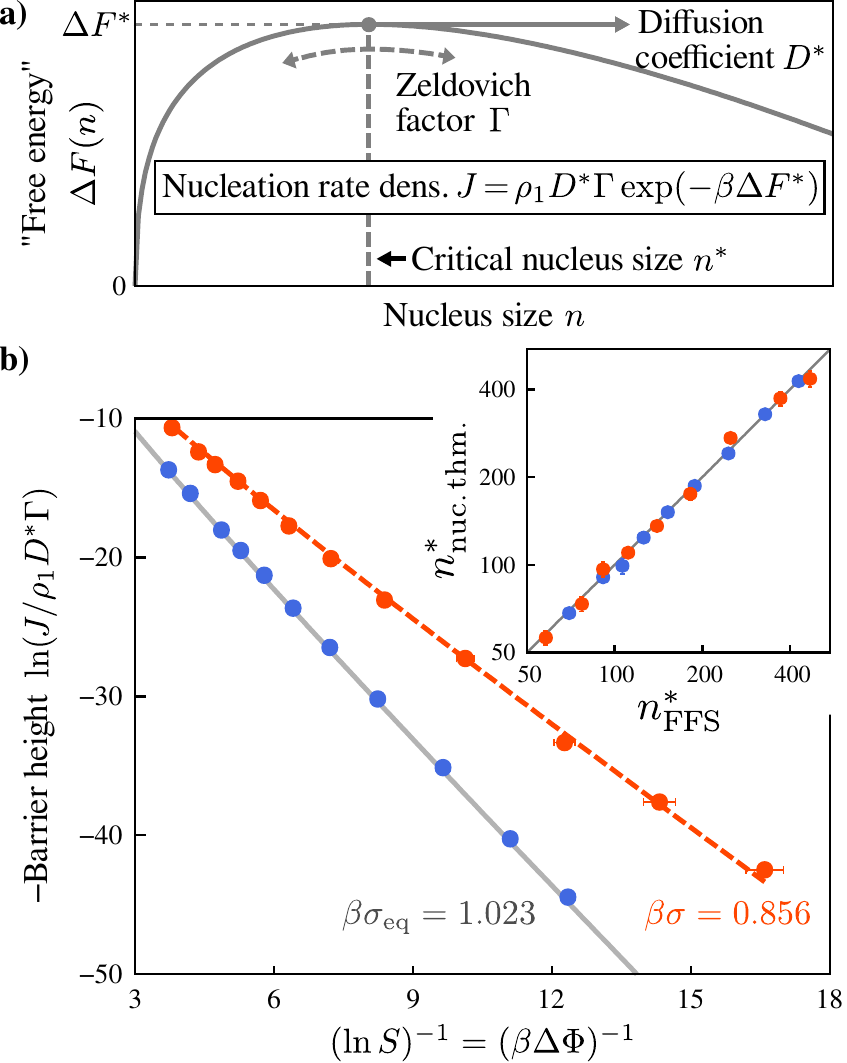}
  \caption{
  Nucleation kinetics at a NESS obey classical nucleation theory (CNT) with modified interfacial properties.
  (a) A schematic illustration of diffusion on an (equilibrium) free-energy landscape, $F(n)$.
  (b) Tests of CNT and the nucleation theorem (inset) for nonequilibrium homogeneous (blue) and inhomogeneous (orange) models under far-from-equilibrium conditions (at $\beta\Delta\mu_{\text{coex}} = 1.87$ using the same parameters as Figs. \hyperref[fig:1]{1(c)} and \hyperref[fig:1]{1(d)}.
  Solid and dashed curves show the equilibrium prediction and a fit of the inhomogeneous results to the CNT rate equation, respectively.}
  \label{fig:2}
\end{figure}

We can now address the central question of this work: To what extent can equilibrium descriptions of nucleation be applied to phase separation at a NESS?
The most widely used theoretical framework for describing nucleation in systems ranging from atomic and molecular fluids to colloidal and biomolecular materials is classical nucleation theory (CNT)~\cite{oxtoby1992homogeneous,sear2007nucleation}.
In its most general form, equilibrium CNT predicts that nucleation follows a minimum free-energy pathway along a reaction coordinate corresponding to the size of a nucleus of the stable phase.
This pathway crosses a free-energy barrier that arises from the competition between the lower thermodynamic potential of the stable phase and the positive interfacial free energy between the nucleus and the bulk metastable phase.
CNT predicts that the homogeneous nucleation rate density is the product of a prefactor and a Boltzmann factor corresponding to the height of the barrier, $\Delta F^*$; the prefactor is the product of the monomer number density, $\rho_1$; the speed along the reaction coordinate at the top of the barrier, $D^*$; and the Zeldovich factor, $\Gamma$, that accounts for fluctuations that cross the barrier but return to the metastable state [\figref{2}{a}].
After taking into account the interfacial free energy due to the macroscopic line tension and microscopic nucleus size fluctuations, CNT has been shown to provide a quantitative description of nucleation in the two-dimensional equilibrium lattice gas model~\cite{ryu2010validity}.

We employ forward-flux sampling (FFS)~\cite{allen2009forward} to compute the nucleation rate density, $J$, and the commitment probability to the stable phase, $\phi(n)$, using the largest nucleus size, $n$, as the reaction coordinate.
The critical nucleus size, $n^*$, is found where $\phi(n^*) = \nicefrac{1}{2}$~\cite{hummer2004}, and the Zeldovich factor can be calculated by fitting $\phi(n)$ to an approximately harmonic barrier in the vicinity of $n^*$.
We also independently measure the number density of bonding-state monomers in the vapor phase, $\rho_1$, and the diffusion coefficient, $D^*$, from nucleus-size fluctuations near $n^*$~\cite{auer2004quantitative}.
We are therefore able to isolate the factor in the CNT rate equation that pertains to the (nonequilibrium) nucleation barrier by computing $\ln (J / \rho_1 D^* \Gamma)$ as a function of the supersaturation, $S \equiv \exp(\beta\Delta\Phi)$, which we control by tuning $\Delta\mu$ (Appendix~\hyperref[app:noneq-line-tension]{B}).

We first test the prediction of the fundamental nucleation theorem, $n^* = -\partial \ln (J / \rho_1 D^* \Gamma) / \partial \ln S + 1$, for nucleating a stable liquid phase from a supersaturated vapor phase~\cite{sear2007nucleation}.
This prediction holds as long as the interfacial free energy is independent of the supersaturation, regardless of the functional form of the nucleation barrier.
The results of representative simulations shown in the inset of \figref{2}{b} demonstrate excellent agreement between the critical nucleus sizes obtained from FFS, $n^*_\text{FFS}$, and the sizes inferred from this theorem, $n^*_\text{nuc.\,thm.}$.
This provides evidence that the fundamental premise of CNT---namely, that the rate-limiting step coincides with the formation of a critical nucleus of the stable bulk phase---applies to nucleation at a NESS in the regime $\beta\Delta\Phi \lesssim 1$.

However, when examining the supersaturation dependence of the apparent nucleation barrier [\figref{2}{b}], we discover a surprising deviation from the equilibrium lattice gas: Although the interfacial contribution still scales with the perimeter of the two-dimensional nucleus, the inferred line tension, $\sigma$, differs from the equilibrium value, $\sigma_{\text{eq}}$.
This deviation only occurs in the case of inhomogeneous reactions, which can be seen by comparing the homogeneous and inhomogeneous results with the equilibrium barrier height in \figref{2}{b}.
These observations indicate that CNT can be extended to describe phase separation at a NESS, but that the nucleation rate can differ by orders of magnitude from predictions based on equilibrium interfacial properties in the case of nonequilibrium inhomogeneous reactions.

\begin{figure}
  \includegraphics[width=\columnwidth]{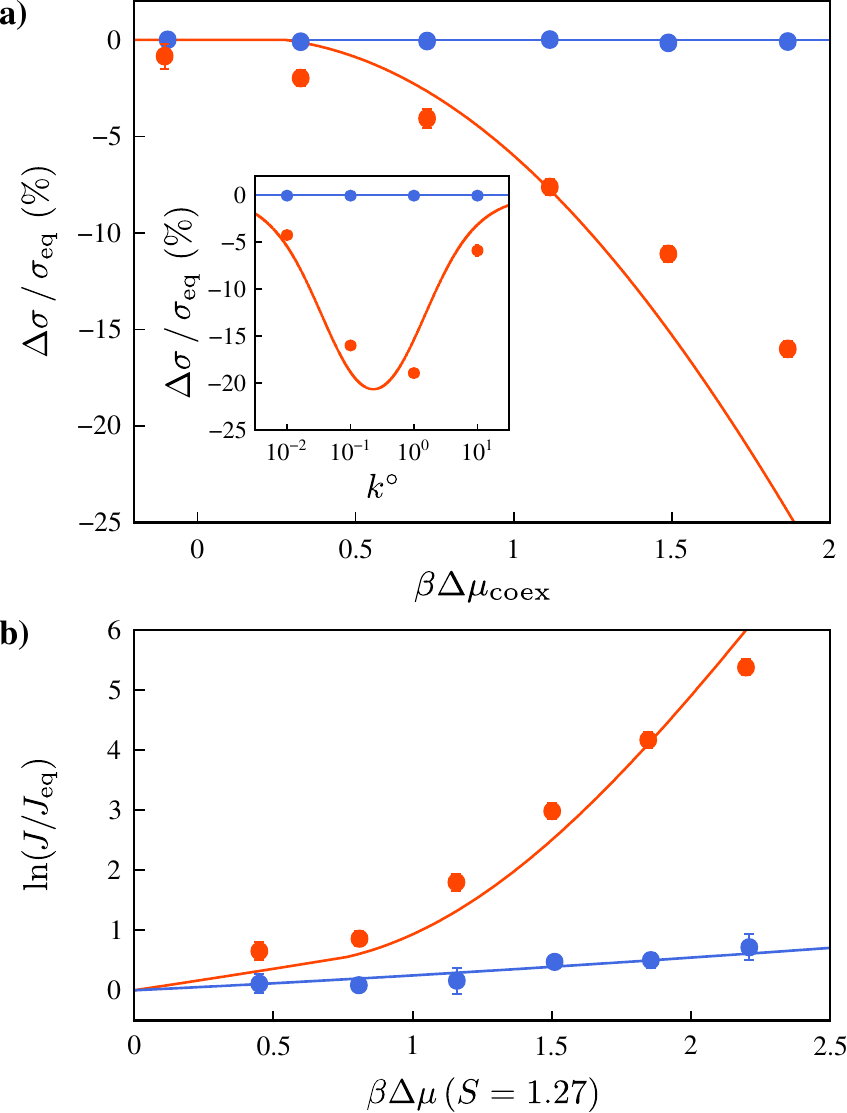}
  \caption{
  Inhomogeneous reactions at a NESS alter the interfacial tension, which strongly affects the nucleation kinetics.
    (a) Deviation of the nonequilibrium line tension, $\Delta\sigma\equiv\sigma - \sigma_\text{eq}$, as determined from nucleation rate calculations, with respect to $\beta\Delta\mu_{\text{coex}}$ and $k^\circ$ (inset).
    The simulation parameters are the same as in Figs. \hyperref[fig:1]{1(c)} and \hyperref[fig:1]{1(d)}.
  (b) Comparison of nonequilibrium nucleation rate densities, $J$, to corresponding equilibrium rate densities, $J_\text{eq}$, at constant supersaturation, $S=1.27$ (see text).
  Orange and blue colors indicate nonequilibrium inhomogeneous and homogeneous models, respectively.
  Symbols report FFS results, and lines show theoretical predictions.
  }
  \label{fig:3}
\end{figure}

Our simulations reveal that the liquid--vapor interfacial properties are influenced by how far the system is driven out of equilibrium.
To illustrate this effect, we perform simulations corresponding to an isothermal experiment in which the total number of particles is conserved, such that $\beta\epsilon$ and $\rho_{\textit{v}}$ are held constant.
We find that the line tension for inhomogeneous reactions deviates farther from the equilibrium value as we increase $\beta\Delta\mu$ at coexistence [\figref{3}{a}].
We also find that the interfacial properties depend on the relative timescale, $k^\circ$, between $\textit{I}\rightleftharpoons\textit{B}$ transitions and the rate of particle attachment to the nucleus, normalized by its perimeter.
To test the sensitivity of the line tension to the ratio of these timescales, we calculate the line tension deviation, $\Delta\sigma$, while holding $\Delta f_{\text{res}} + \Delta\mu$ constant [inset of \figref{3}{a}].
We find that $\Delta\sigma$ is nonzero over a wide range of $k^\circ$, with the greatest deviation occurring when these timescales are comparable ($k^\circ\approx 1$).
However, we recover the equilibrium line tension in the limit of either zero reactive flux ($k^\circ \rightarrow0$) or infinitely fast reactions ($k^\circ \rightarrow\infty$), as the system reverts either to a true equilibrium or to a NESS in which $\Delta\Delta f \rightarrow 0$, respectively.

We can understand these results by considering a theoretical model that captures the qualitative behavior of the nonequilibrium interface.
We make the approximation that particle exchange between the open system and the reservoir relaxes to the steady-state distribution more quickly than the local environment around a particle changes.
Within this ``Fixed Local Environment approXimation'' (FLEX), the steady-state number densities $\pI$ and $\pB$ map to an \textit{effective} equilibrium system with fugacities $\zIT$ and $\zBT$ (Appendix~\hyperref[app:FLEX]{C}).
Examining the internal free-energy difference $\beta\Delta f \equiv -\ln(\zBT/\zIT)$ within the FLEX framework shows that a common effective equilibrium describes both phases if $\kIB$ is constant, corresponding to homogeneous reactions, regardless of $\Delta\mu$.
Conversely, a different effective equilibrium is needed for each phase if $\kIB$ depends on $u$, corresponding to inhomogeneous reactions.

To predict the nonequilibrium interfacial tension from FLEX, we employ a solid-on-solid model~\cite{saito1996statistical} of an interface at coexistence.
We first find $\beta\Delta\mu_{\text{coex}}$ by setting $S_{\text{FLEX}} \equiv \left[\pB / (1 - \pB)\right]_{u=2\epsilon} = 1$, where the fixed local environment $u=2\epsilon$ is assumed based on the particle--hole symmetry of the equilibrium lattice gas.
We then calculate the effective energy of attaching a single bonding-state adatom to a flat interface, $\beta\tilde\epsilon \equiv \ln\left[ \pB / (1-\pB) \right]_{u=\epsilon}$, at the coexistence points $\beta\Delta\mu_{\text{coex}}$ (Appendix~\hyperref[app:FLEX-interface]{D}).
Importantly, $\tilde\epsilon$ only differs from $\epsilon$ with nonequilibrium inhomogeneous reactions.
Finally, we estimate the nonequilibrium line tension by evaluating an equilibrium expression for $\sigma(\beta\tilde\epsilon)$~\cite{shneidman1999analytical} [solid curves in \figref{3}{a}].
In our inhomogeneous simulations, $\kIB$ is a decreasing function of $u$,
leading to a lower $\Delta f$ and thus a higher population of bonding-state particles at the interface than would be expected based on the effective equilibrium model of the bulk liquid phase.
This enrichment of bonding-state particles at the interface relative to the liquid phase reduces the effective adatom bonding energy in our theory, such that $|\tilde\epsilon| \le |\epsilon|$, and lowers the effective free-energy cost of the interface.

Our key insight from this theory is that nonequilibrium interfacial properties emerge when the bulk phases and the liquid--vapor interface are described by \textit{different} effective equilibrium models.
Consequently, when $\tilde\epsilon \ne \epsilon$, bonding particles attached to the interface of a critical nucleus may be attracted either more or less strongly, per nearest-neighbor interaction, than in the bulk liquid phase.
Our theory captures both the sign and the approximate functional form of $\Delta\sigma$ with respect to $\beta\Delta\mu_{\text{coex}}$, as well as the nonmonotonic dependence of $\Delta\sigma$ on the relative reaction timescale $k^\circ$.
While the precise form of $\Delta\sigma$ depends on our choice of simulation parameters, the generality of our theory suggests that a nonzero $\Delta\sigma$ can arise whenever the effective internal free-energy difference is a function of the local environment.

Finally, to highlight the control over nucleation rates imparted by inhomogeneous reactions, we compare the nonequilibrium nucleation rate to that of an equilibrium fluid with the same $\beta\epsilon$, $S$, and $\rho_\textit{v}$ [\figref{3}{b}].
In agreement with our theory, our simulations show that the nucleation rate can be increased by orders of magnitude relative to the corresponding equilibrium system by driving the fluid far from equilibrium ($\beta\Delta\mu \gg 1$).
The magnitude of this effect is far greater in the inhomogeneous than in the homogeneous model due to the dominant role of the line tension in determining the nucleation barrier, and thus the nucleation rate.
Inhomogeneous reactions can therefore break the usual relationship between high supersaturation and fast nucleation, offering a novel way to control nucleation kinetics in nonequilibrium fluids.

In conclusion, we have introduced a strategy for simulating nonequilibrium phase transformations within the framework of stochastic thermodynamics.
By showing that inhomogeneous chemical reactions can give rise to nonequilibrium interfacial tensions, our work reveals a mechanism for decoupling nucleation rates from thermodynamic driving forces at a NESS.
Our findings provide further evidence~\cite{berry2018physical,weber2019physics} that nonequilibrium phase transformations may follow the same phenomenological laws as equilibrium systems under rather general conditions.
Detecting nonequilibrium effects may thus require careful measurements of interfacial material properties.

We emphasize that our qualitative results do not depend on the specific form of the reaction rates: The only essential ingredient is an inhomogeneously driven reaction that is either promoted or suppressed by variations in the local potential energy.
For example, if the reactive flux through the driven pathway is enhanced at low potential energies, as in our simulations, then our model can describe either preferentially driven deactivation ($\textit{B}\rightarrow\textit{I}$) in the liquid phase or preferentially driven activation ($\textit{I}\rightarrow\textit{B}$) in the vapor phase.
The former scenario represents an implicit description of enzyme-mediated deactivation, in which the chemical fuel is uniformly distributed but the enzymes that catalyze $\textit{B}\rightarrow\textit{I}$ reactions preferentially partition into the condensed, low-potential-energy phase.
Such a scenario has been proposed to describe inhomogeneous enzyme distributions associated with stress granules and other biological condensates~\cite{soding2020mechanisms,hondele2020membraneless,oflynn2021role}.

Our results are applicable to a range of experimental systems broadly described as living or active.
Our prediction of a nonequilibrium surface tension could be tested in the context of intracellular condensates using light-activated corelets~\cite{bracha2018mapping}, which have recently been applied to study condensate nucleation \textit{in vivo}~\cite{shimobayashi2021nucleation}.
Our model could also be applied to synthetic active polypeptide coacervates~\cite{nakashima2018reversible,spaeth2021molecular,nakashima2021active} or DNA liquids~\cite{saleh2020enzymatic} in which the association/hybridization reactions are engineered to respond to energy input in a manner that is dependent on the local protein/DNA concentration.
In both contexts, our results suggest a road map for controlling self-assembly kinetics far from thermal equilibrium.

This work is supported by the National Science Foundation (DMR-2143670).

\clearpage


\textit{Appendix A: Nonequilibrium lattice-gas model.}---
\label{app:model}
We extend the two-dimensional square lattice-gas model by incorporating two particle internal states: a bonding state (\textit{B}) and an inert state (\textit{I}).
\textit{B}-state particles interact with nearest-neighbor B-state particles with bonding strength $\epsilon < 0$.
By contrast, \textit{I}-state particles are isoenergetic to empty lattice sites and thus do not interact with nearest-neighbor particles.
Here we consider an open system in contact with a particle reservoir, such that \textit{B} and \textit{I}-state particles have fugacities $\zB$ and $\zI$, respectively, in the reservoir.
Open systems have similar advantages for studying nonequilibrium phase transitions as the grand-canonical ensemble does for equilibrium systems, including the elimination of interfaces and a resulting reduction of finite size effects~\cite{wilding1995critical}.

\begin{figure}[!h]
\includegraphics[width=0.4\columnwidth]{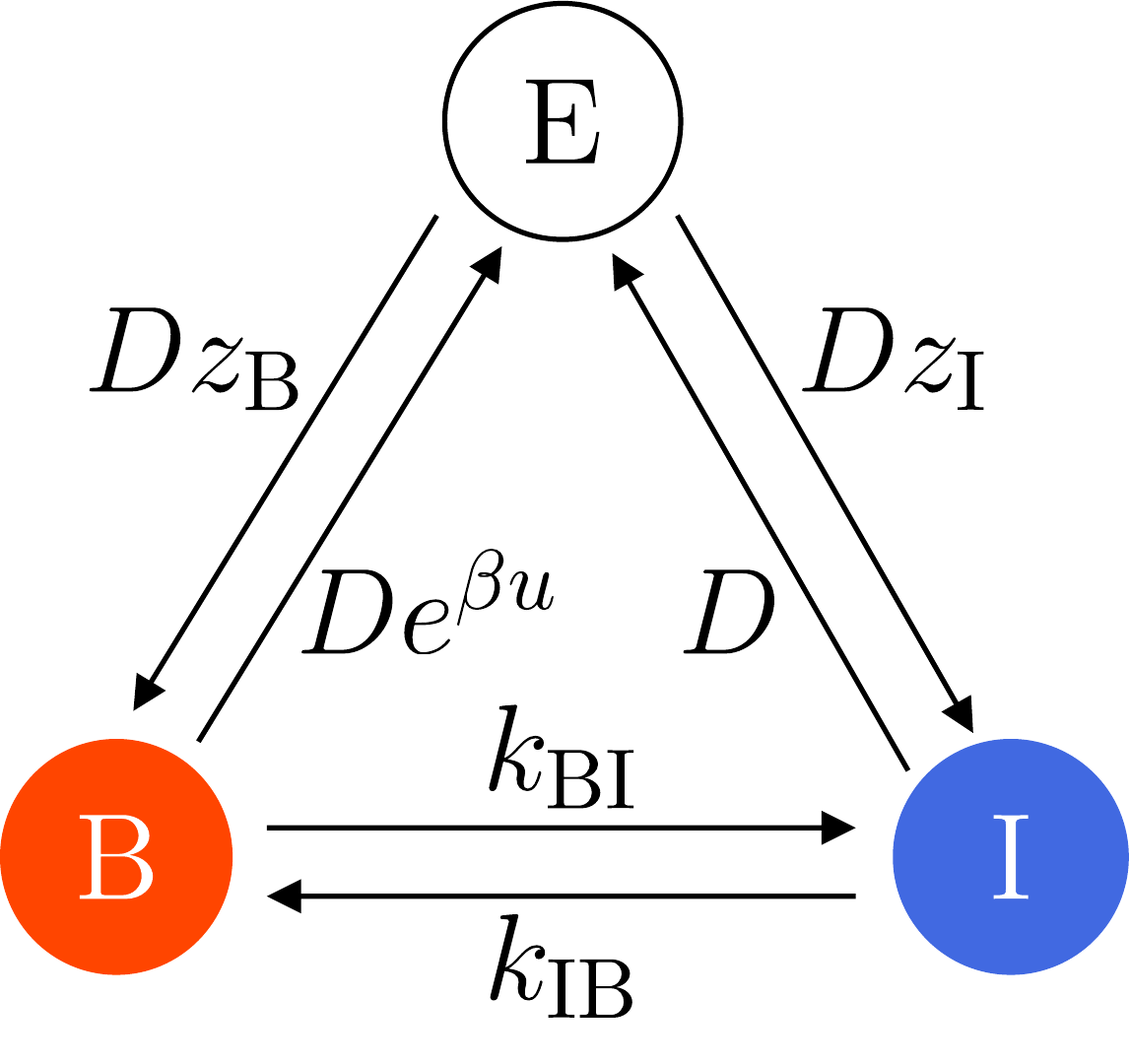}\vskip-1.5ex
\caption{
Kinetic scheme of particle exchange and internal chemical reactions.
In our simulations of an open system, each lattice site stochastically transitions between being unoccupied (\textit{E}) or being occupied by either a bonding~(\textit{B}) or inert~(\textit{I}) particle with the specified transition rates.}
\label{fig:4}
\end{figure}

Utilizing the framework of stochastic thermodynamics, we model the kinetics of particle insertion, removal, and reactions between internal states using Markovian transitions that obey local detailed balance~\cite{seifert2012stochastic,van2015ensemble}.
Particle insertion and removal rates depend on the reservoir fugacities, $\zB$ and $\zI$; the local potential energy $u$ due to nearest-neighbor interactions at a particular lattice site; and the base exchange rate, $D$, between the open system and the reservoir (Fig.~\ref{fig:4}).
Reactions between the \textit{B} and \textit{I} states occur with forward and backward rates $k_{\textit{BI}}$ and $k_{\textit{IB}}$.
We introduce dimensionless ratios between tje reaction and particle-exchange rates, $\kBI \equiv D^{-1}k_{\textit{BI}}$ and $\kIB \equiv D^{-1}k_{\textit{IB}}$, for notational simplicity.
We simulate the stochastic evolution of the system via the kinetic Monte Carlo method~\cite{gillespie2007stochastic}.

We define the nonequilibrium drive $\Delta\mu$ along the single-cycle network (Fig.~\ref{fig:4}) in the \textit{B}-to-\textit{I} direction,
\begin{equation}
  \label{eq:Dmu}
  \beta\Delta\mu = \ln\left[{\zB \kBI}/{\zI\kIB e^{\beta u}}\right]. \tag{A1}
\end{equation}
Time-reversal symmetry is broken when the system is driven out-of-equilibrium ($\Delta\mu \ne 0$), resulting in a nonzero net probability current.
Rearranging \eqref{eq:Dmu} gives the local detailed balance condition for $\textit{I}\rightleftharpoons\textit{B}$ reactions in terms of the chemical drive $\Delta\mu$,
\begin{equation}
  \label{eq:local-detailed-balance}
  {\kBI}/{\kIB} = \exp(\beta u + \beta\Delta f_{\text{res}} + \beta\Delta\mu),
  \tag{A2}
\end{equation}
where $\beta\Delta f_\text{res} \equiv -\ln(\zB/\zI)$ is the internal free-energy difference in the reservoir.

In our simulations, we consider two specific choices for the backward reaction rate, $\kIB$, in order to model homogeneous and inhomogeneous chemical reactions.
For \textit{homogeneous} systems, we set $\kIB$ equal to a constant $k^\circ$ representing the ratio between the timescales for chemical reactions and particle transport.
For \textit{inhomogeneous} systems, we assume that $\kIB$ is $u$-dependent and takes a Metropolis form, $\kIB = {k^\circ \min[1, \; \exp(-\beta u - \beta\Delta f_{\text{res}} - \beta\Delta\mu)]}$.
Note that $\kBI$ follows from the local detailed balance condition, \eqref{eq:local-detailed-balance}, in both cases.

We quantify the extent of inhomogeneous chemical reactions by estimating the effective free-energy difference between the two particle internal states, $\Delta f$, from simulations of each bulk phase.
A lattice configuration is defined by the lattice-site occupancies, $\{c(\bm{r})\}$, where $c\in\{\textit{E},\textit{B},\textit{I}\}$ and $z_{\textit{E}} = 1$.
The equilibrium probability that the tagged site at the origin, $\bm{r}=0$, is in state $i$ is
\begin{gather}
  \label{eq:semigrand-bennet-appendix}
  \frac{p_{i(\bm{r}=0)}^{\text{eq}}}{p_{j(\bm{r}=0)}^{\text{eq}}} = \left(\frac{z_i}{z_j}\right)\! \left\langle e^{-\beta \sum_{\bm{r'}} u[i,c(\bm{r'})] - u[j,c(\bm{r'})]} \right\rangle_{\!j(\bm{r}=0)}\!, \tag{A3}
\end{gather}
where summation is over the nearest-neighbor sites $\bm{r'}$ of the tagged site, $u[i,j]$ is the potential energy between nearest-neighbor lattice sites in states $i$ and $j$, and angle brackets indicate an ensemble average conditioned on the tagged site being in the indicated state.
We use \eqref{eq:semigrand-bennet-appendix} to define the effective $\Delta f$ by substituting $p^{\text{eq}}$ with the NESS distribution, $p$, and averaging over the NESS,
\begin{equation}
  \label{eq:Df}
  \beta\Delta f = -\ln \left( \frac{p_{\textit{B}}}{p_{\textit{I}}} \right) + \ln \left\langle e^{-\beta \sum_{\bm{r'}} u[\textit{B},c(\bm{r'})]} \right\rangle_{\!\textit{I}(\bm{r}=0)}\!. \tag{A4}
\end{equation}

\vspace{1em}
\textit{Appendix B: Determining the interfacial tension in nonequilibrium nucleation simulations.}---
\label{app:noneq-line-tension}
The free-energy landscape along the nucleus-size reaction coordinate, $n$, in the equilibrium lattice gas is~\cite{ryu2010validity}
\begin{equation}
  \label{eq:CNT_landscape}
  \beta F(n) = \beta\sigma\sqrt{4\pi n} - \beta\Delta\Phi n + (5/4)\ln n + d, \tag{B1}
\end{equation}
where $d$ is a constant chosen to equate the \textit{B}-state monomer number density in the vapor phase, $\rho_1$, and $\exp[-\beta F(1)]$, such that the barrier height is ${\Delta F^* \equiv F(n^*) - F(1)}$.
From the CNT rate density, ${J = \rho_1 D^* \Gamma \exp(-\beta F^*)}$, we obtain
\begin{align*}
  \label{eq:CNT_rate}
  \ln\left(\dfrac{J}{\rho_1D^*\Gamma}\right) =& \: \beta\Delta\Phi(n^*-1) - \beta\sigma\sqrt{4\pi}\left(\sqrt{n^*}-1\right) \\ &- (5/4)\ln n^*\!, \tag{B2}
\end{align*}
where $n^* = 25/(-\beta\sigma\sqrt{4\pi} + \sqrt{4\pi\beta^2\sigma^2 + 20\beta\Delta\Phi})^2$ is the critical nucleus size, and $\ln (J/\rho_1 D^* \Gamma)$ is an approximately linear function of $1/\beta\Delta\Phi$ with slope proportional to $-\sigma^2$ [\figref{2}{b}].
Using FFS simulations on a $64\times64$ lattice, we measure $\rho_1$ in the vapor phase and calculate $D^*$ by analyzing the diffusive behavior of the reaction coordinate when $n \approx n^*$.
The Zeldovich factor, $\Gamma$, is found independently by fitting the commitment probabilities, $\phi(n)$, calculated in FFS simulations,
\begin{equation}
  \label{eq:zeldovich}
  \phi(n) \approx \frac{1}{2} \text{erf}\left[\Gamma \sqrt{\pi} (n - n^*)\right] + \frac{1}{2}, \tag{B3}
\end{equation}
where erf is the error function, and we have assumed that the landscape is approximately parabolic near $n \approx n^*$.
We obtain the line tension, $\sigma$, by fitting \eqref{eq:CNT_rate} over a range of $\beta\Delta\Phi$ values determined from nonequilibrium umbrella sampling, using $\sigma$ as the sole fitting parameter.

\begin{figure}[t]
\includegraphics[width=0.9\columnwidth]{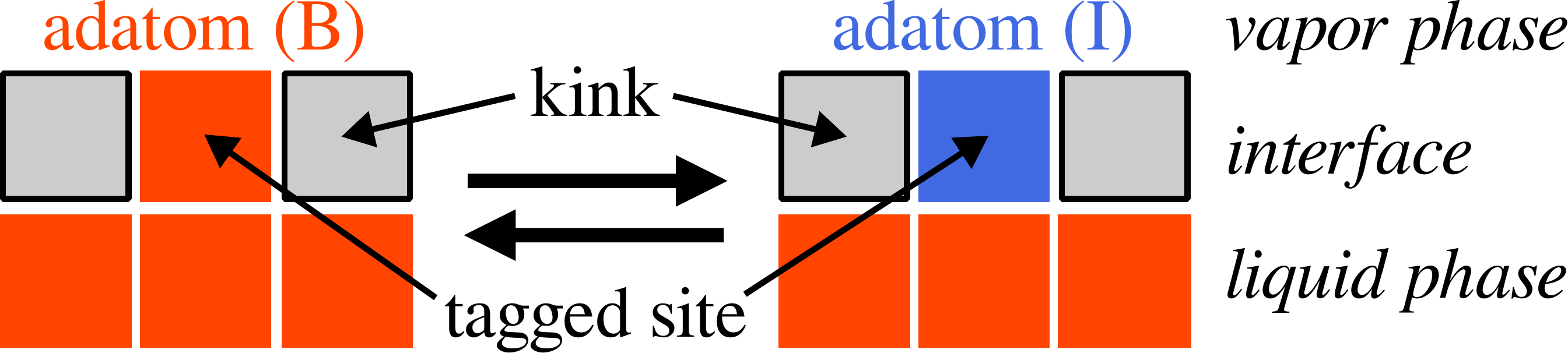}
\caption{
FLEX schematic of a single-layer configuration at a liquid--vapor interface.
The effective bonding energy at the interface is obtained from the steady-state distribution at the tagged site under a fixed local configuration.
Colors correspond to the same lattice-site states as in Fig.~\ref{fig:4}.}
\label{fig:5}
\end{figure}

\vspace{1.5em}
\textit{Appendix C: Fixed Local Environment approXimation (FLEX).}---
\label{app:FLEX}
In the Fixed Local Environment approXimation (FLEX), we assume that particle exchange between the open system and the reservoir relaxes to the steady state more rapidly than any change in the local configuration, or environment, around a tagged lattice site.
Specifically, we represent the configuration around a tagged lattice site by a fixed number of nearest-neighbor \textit{B}-state particles (Fig.~\ref{fig:5}).
We then calculate the single-site steady-state distribution, $\tilde{\rho}_i$, from the Markovian transition network shown in Fig.~\ref{fig:4}; $\tilde{\rho}_i$ may be regarded as the number density of a particle, if $i=$ \textit{B} or \textit{I}, or a vacancy, if $i=$ \textit{E}.

We map our nonequilibrium model to an \textit{effective} equilibrium that has the same steady-state distribution $\tilde{\rho}$ as that predicted by FLEX.
To this end, we define effective fugacities in the open system, $\zBT \equiv (\pB/\pE)\exp(\beta u)$ and $\zIT \equiv \pI/\pE$, and the single-site partition function $\tilde{\xi}=1+\zBT+\zIT$.
Depending on the functional form of $\kIB$, the effective fugacities may depend on $u$, and the liquid and vapor phases may be mapped to different effective equilibrium models.
We therefore calculate the effective internal free-energy difference, $\Delta f$, between the \textit{B} and \textit{I} states in the open system, as in \eqref{eq:Df}.
Within FLEX, $\beta\Delta f \equiv -\ln(\zBT/\zIT)$ is related to $\beta\Delta f_\text{res}$ by
\begin{equation}
  \beta\Delta f = \beta\Delta f_\text{res} + \ln\left[\dfrac{1+\kIB(1+e^{\beta\Delta f_\text{res}})e^{\beta\Delta\mu}}{1+\kIB(1+e^{\beta\Delta f_\text{res}})}\right].
  \label{eq:FLEX_inhomogeneity} \tag{C1}
\end{equation}
\eqref{eq:FLEX_inhomogeneity} predicts the requirements for coexisting phases, which have different average potential energies per lattice site, to be thermodynamically inhomogeneous: For $\Delta f$ to be $u$-dependent, $\kIB$ must be $u$-dependent and chemical drive must be applied ($\Delta\mu\ne0$).
These conditions are consistent with the simulation results shown in \figref{1}{d}.

\vspace{1.5em}
\textit{Appendix D: FLEX prediction of the nonequilibrium interfacial tension.}---
\label{app:FLEX-interface}
Phase coexistence in the two-dimensional equilibrium lattice gas occurs at $\mu = 2\epsilon$, where $\mu$ is the particle chemical potential, due to particle--hole symmetry \cite{pathria1996statistical}.
The resulting supersaturation $S \approx \exp[\beta(\mu - 2\epsilon)]$ in the equilibrium model can be interpreted as the ratio $\rho / (1 - \rho)$ at a tagged lattice site with exactly two neighboring particles, where $\rho$ is the particle number density.
Assuming that particle--hole symmetry is a reasonable approximation for the effective equilibrium as well, we define the FLEX supersaturation, $S_{\text{FLEX}}$, based on the steady-state distribution at a tagged lattice site with $u=2\epsilon$,
\begin{equation}
    S_\text{FLEX} \equiv \left[\dfrac{\pB}{1-\pB}\right]_{u=2\epsilon}
    \!= \dfrac{\zBT(2\epsilon;\Delta\mu) e^{-2\beta\epsilon}}{1+\zIT(2\epsilon;\Delta\mu)}. \tag{D1}
    \label{eq:S_FLEX}
\end{equation}
We then predict the NESS coexistence point by solving for the chemical drive at which $S_{\text{FLEX}}=1$, subject to an imposed total particle density in the vapor phase.

We use FLEX to predict the interfacial tension by considering the attachment of a single bonding-state particle to a flat liquid--vapor interface in a solid-on-solid model at phase coexistence.
We focus on the effective bonding energy $\beta\tilde{\epsilon}$ of a single adatom, since the coexistence condition $S_{\text{FLEX}} = 1$ implies that the formation of a kink on the interface (Fig.~\ref{fig:5}) incurs no (effective) free-energy cost.
To determine $\beta\tilde{\epsilon}$, we apply FLEX to a tagged adatom site with $u=\epsilon$ at the predicted coexistence point, $\Delta\mu_{\text{coex}}$.
Because the equilibrium free-energy cost to attach an adatom to a flat interface is $\beta\epsilon$, we define the effective bonding energy $\beta\tilde{\epsilon}$ in the same way:
\begin{equation}
    \label{eq:effective_bond_strength}
    \beta\tilde{\epsilon} \equiv \ln\left[\dfrac{\pB}{1-\pB}\right]_{u=\epsilon}
    \!= \ln \left[\dfrac{\zBT(\epsilon; \Delta\mu_\text{coex})}{1\!+\!\zIT(\epsilon; \Delta\mu_\text{coex})}\right]-\beta\epsilon. \tag{D2}
\end{equation}
In homogeneous systems, this prediction reduces to $\beta\tilde{\epsilon} = \beta\epsilon$, meaning that the effective adatom interaction strength does not change no matter how far the system is driven out of equilibrium.
However, in the case of inhomogeneous chemical reactions, $\beta\tilde{\epsilon}$ may differ from $\beta\epsilon$.

Finally, we predict the nonequilibrium interfacial tension, $\sigma$, using the adatom interaction strength $\beta\tilde{\epsilon}$ at the interface and the equilibrium formula \cite{shneidman1999analytical}
\begin{equation}
\label{eq:sigma_analytical}
    \sigma(\tilde\epsilon) = \!\sqrt{\dfrac{4\tilde\epsilon\beta^{-2}}{\pi\chi(\beta)}\!\!\int_{\beta_c}^\beta \!\!\! K'\!\left(\!\dfrac{8[\cosh(\beta'\tilde\epsilon)\!-\!1]}{[\cosh(\beta'\tilde\epsilon)\!+\!1]^2}\!\right) \!\!\left[\!\dfrac{\cosh(\beta'\tilde\epsilon)\!-\!3}{\sinh(\beta'\tilde\epsilon)}\!\right] \!d\beta'}\!, \tag{D3} \vspace{1em}
\end{equation}
where $K'$ is the elliptic integral of the first kind, $\chi(\beta) = [1-\sinh^{-4}({\beta\tilde{\epsilon}/2})]^{1/8}$, and $\beta_\text{c}$ is the inverse critical temperature given by $\beta_\text{c}|\tilde{\epsilon}| = 2\ln(1+\sqrt{2})$.
We find that this FLEX prediction qualitatively explains the decreasing trend of the line tension with respect to the nonequilibrium drive in the inhomogeneous model [see \figref{3}{a}].

\nocite{noya2008}
\nocite{hansen2013simpleliquids}

\clearpage


%

\renewcommand{\theequation}{S\arabic{equation}}
\renewcommand{\thefigure}{S\arabic{figure}}

\setcounter{figure}{0}
\renewcommand{\theequation}{S\arabic{equation}}
\renewcommand{\thefigure}{S\arabic{figure}}

\clearpage

\onecolumngrid

\section*{Supplementary Information for ``Tuning nucleation kinetics via nonequilibrium chemical reactions''}

\section{Nonequilibrium phase coexistence simulations}

\subsection{Nonequilibrium Umbrella Sampling (NEUS)}
\label{sec:NEUS}

We use a form of nonequilibrium umbrella sampling (NEUS) \cite{warmflash2007umbrella} to obtain the steady-state distribution of the bonding-state particle density, $\rhoB$.
For this purpose, we divide the entire range of $\rhoB$ into non-overlapping boxes and focus on the transition flux between the boxes at steady state.
Importantly, transitions are only possible between adjacent boxes in our simulations because each kinetic Monte-Carlo move can insert or remove at most one bonding-state particle.
Detailed balance between the boxes always holds under this restriction, regardless of whether the system is driven out of equilibrium, as long as the system is at steady state.
The left (\textit{L}) and right (\textit{R}) box-boundary crossing fluxes $\fL(b)$ and $\fR(b)$ therefore satisfy $\fR(b) = \fL(b+1)$ and $\fL(b) = \fR(b-1)$ for each box index $b$.
(We write the box indices only as necessary in what follows.)
The fluxes can be related to the transition probabilities $\{p_{ij}\}$ of reaching the $j$-side boundary starting from the $i$-side boundary of each box, where $i,j\in\{\text{\textit{L},\textit{R}}\}$.
By definition, $p_\textit{LL} + \pLR = \pRL + p_\textit{RR} = 1$.

\begin{figure}[!h]
    \centering
    \includegraphics[scale=0.8]{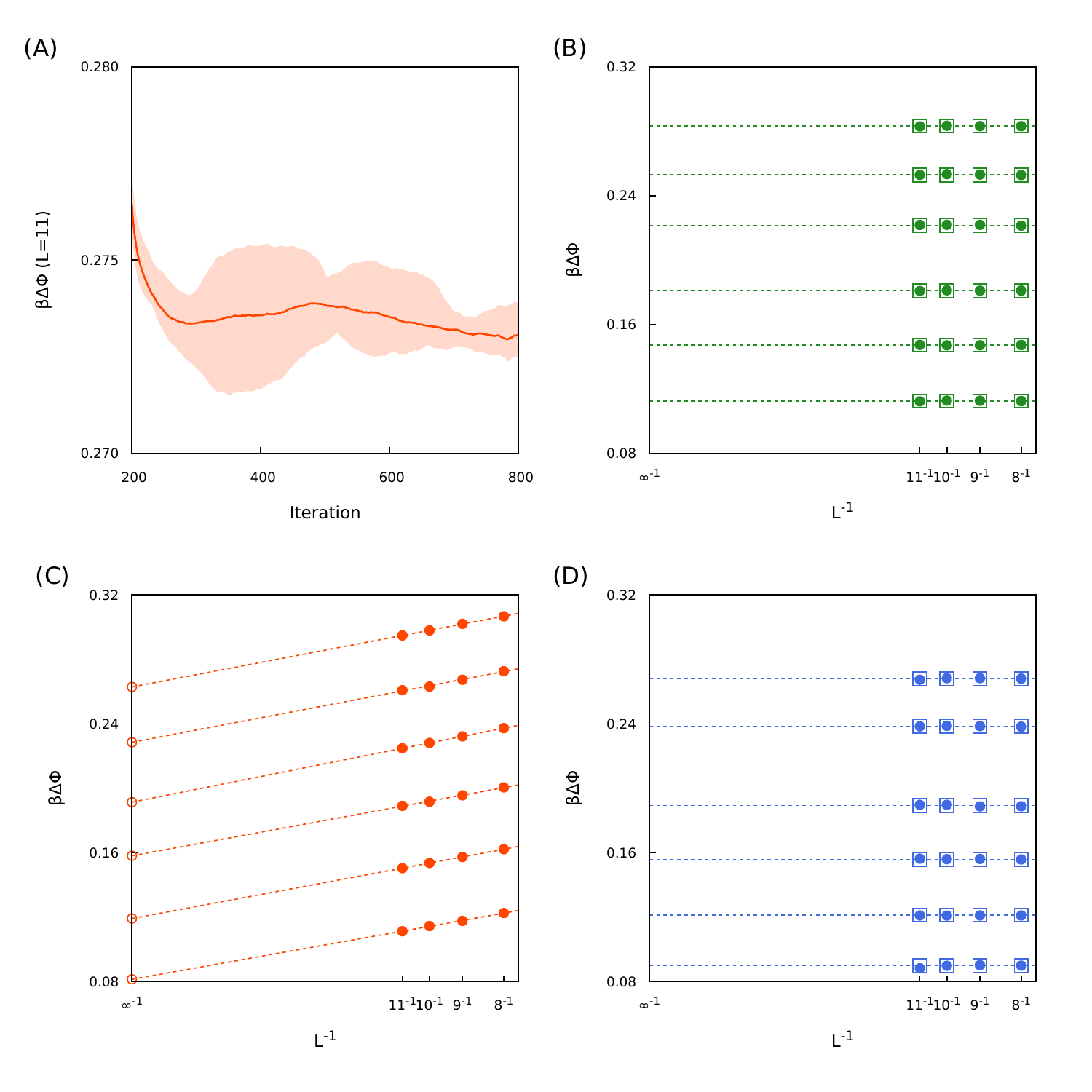}\vskip-3.5ex
    \caption{Representative results of Nonequilibrium Umbrella Sampling.
    (A) Relaxation to the steady state. The value of $\beta\Delta\Phi$ at each iteration is calculated from the average steady-state distribution over the previous 200 iterations. The solid line and the shaded region show the average and the range of $\beta\Delta\Phi$ observed among four independent NEUS trials, respectively.
    (B) Test of NEUS (circles) against a third-order cluster expansion (open squares) for equilibrium systems.
    (C) Analysis of the finite-size effect for nonequilibrium inhomogeneous and (D) homogeneous systems whose coexistence conditions are $\beta\Delta\mu_{\text{coex}} = 1.87$.
    The open squares in (D) are from cluster expansion up to third order in fugacities for the corresponding effective equilibrium systems specified by $\zBT$ and $\zIT$, as defined in \eqref{eq:z_B'} and \eqref{eq:z_I'}, respectively.}
    \label{fig:NEUS}
\end{figure}

Our NEUS algorithm is based on launching trajectories in each box based on the incoming fluxes, and then matching the steady-state distribution between adjacent boxes by enforcing detailed balance.
Equating the incoming and outgoing fluxes at the $i$-side boundary of a box leads to $f_\textit{i} = f_\textit{i} p_\textit{ii} + f_\textit{i'} p_\textit{i'i}$ and $\fL/\fR = \pRL / \pLR$, where $i'$ indicates the opposite side of the box from side $i$.
At each iteration of the algorithm, we use the previously collected ensembles of configurations at the box boundaries and the current estimate of $\{p_\textit{ij}\}$ to launch new trajectories with probabilities $\fL/(\fL+\fR) = \pRL/(\pRL+\pLR)$ and $\fR/(\fL+\fR) = \pLR/(\pRL+\pLR)$ from the left and the right boundaries, respectively, of each box.
We then record the average time, $t(b)$, until each of the launched trajectories exits box $b$; the average time $t(\rhoB;b)$ that a trajectory spends at $\rhoB$ within box $b$; and the probability that a trajectory exits through the $i$-side boundary, $p_i = (p_{\textit{Li}}\fL + p_{\textit{Ri}}\fR)/(\fL+\fR) = p_{i'i}/(\pRL+\pLR)$, of box $b$.
We take $p(\rhoB;b) = t(\rhoB;b)/t(b)$ as the steady-state distribution within box $b$.
The fluxes associated with trajectories originating within box $b$ and exiting via boundary $i$, $g_i(b) \equiv p_i(b) / t(b)$, are related to the overall fluxes by $g_i(b) = f_i(b) w(b)$, where $w(b)$ is the fraction of time that a steady-state trajectory spends in box $b$.
By applying the detailed balance condition between boxes $b$ and $b+1$, $\fR(b) = \fL(b+1)$, we can self-consistently determine the box weights, $w(b+1)/w(b) = g_\text{R}(b)/g_\text{L}(b+1)$, and thus solve for the steady-state distribution over the complete range of $\rhoB$, $p(\rhoB) = p(\rhoB;b)\times w(b)/\sum_b{w(b)}$.

The algorithm is implemented by iteratively obtaining a new ensemble of configurations at each boundary of every box while calculating the steady-state distribution within each box.
The initial configurations at each boundary are sampled from brute-force simulations inside of each box, rejecting any kinetic Monte-Carlo events that would allow the system to cross the box boundaries.
At each subsequent iteration of the algorithm, we first obtain the transition probabilities, $\{p_\textit{ij}\}$, in each box starting from the current ensemble of configurations at the box boundaries.
We then compute the steady-state distribution within each box using trajectories launched from the current ensemble of configurations and the calculated $\{p_\textit{ij}\}$.
We save the configurations from these trajectories that exit the box to form the ensemble for the next iteration of the algorithm.
Finally, we average the steady-state distribution and the transition probabilities within each box over the successive iterations and reconstruct the steady-state distribution over the complete range of $\rhoB$ as described above.

We apply this algorithm to measure the dimensionless thermodynamic driving force between the bulk phases, $\beta\Delta\Phi \equiv L^{-2} \ln \left( p_\textit{l} / p_\textit{v} \right)$, as discussed in the main text.
Representative results of this algorithm are shown in Fig.~\hyperref[fig:NEUS]{S1}.
The quick decay of $\Delta\Phi$ during the initial iterations in Fig.~\hyperref[fig:NEUS]{S1(A)} indicates that the system rapidly relaxes, after which the ensemble of trajectories remains in the steady state.
To verify that NEUS converges to the correct steady-state distribution, we also performed NEUS for an equilibrium lattice gas and confirmed that the results match a cluster expansion \cite{hansen2013simpleliquids} up to the third order in $\zB$ and $\zI$ [Fig.~\hyperref[fig:NEUS]{S1(B)}].

When performing systematic calculations at different lattice sizes, we observe a linear dependence of $\Delta\Phi$ with respect to the inverse of the system size, $L^{-1}$, in the case of nonequilibrium inhomogeneous systems [Figs.~\hyperref[fig:NEUS]{S1(C)} and ~\hyperref[fig:NEUS]{S1(D)}].
By contrast, we do not observe any system-size dependence for the nonequilibrium homogeneous and equilibrium systems (Figs.~\hyperref[fig:NEUS]{S1(B)} and \hyperref[fig:NEUS]{S1(D)}).
Thus, in the case of inhomogeneous systems, we calculate the supersaturation in the thermodynamic limit by extrapolating the values obtained from simulations performed in finite systems to the infinite system size [Fig.~\hyperref[fig:NEUS]{S1(C)}].
We attribute this system-size dependence to the broken particle--hole symmetry between the liquid and the vapor phases induced by the inhomogeneous chemical reactions.
We further note that the steady-state distributions for the nonequilibrium homogeneous systems and their equivalent equilibrium systems are identical [Fig.~\hyperref[fig:NEUS]{S1(D)}], which is consistent with the prediction of an effective equilibrium for homogeneous systems discussed in \secref{sec:inhomogeneous}.

\subsection{Validation of phase coexistence via direct coexistence simulations}

Phase coexistence is established at a NESS when the net flux of particles, the net flux of thermal energy, and the pressure difference between two phases are zero.
These conditions are analogous to the equilibrium phase-coexistence criteria of equal chemical potentials, temperatures, and pressures.
In our model, coupling to a single particle reservoir regardless of which phase is currently occupying the lattice, along with the local detailed balance condition governing the reaction rates, ensures that the net particle and heat fluxes vanish between the two phases.
In order to satisfy mechanical equilibrium, we propose that the nonequilibrium potential difference defined above, $\Delta\Phi$, should be set equal to zero.
This choice is motivated by analogy to equilibrium statistical mechanics, in which case $\Delta\Phi$ is equal to the difference between the grand potential densities of the two phases.
Because the grand potential is proportional to the pressure at equilibrium, setting $\Delta\Phi$ equal to zero guarantees mechanical balance at equilibrium.
Under nonequilibrium conditions, however, the relation between the nonequilibrium grand potential and the pressure does not hold.
Nonetheless, $\Delta\Phi = 0$ still implies that the steady-state probabilities of the two phases occupying a given volume are equal at a NESS.
In our lattice model, this condition also means that a long trajectory spends an equal amount of time with each phase completely occupying the lattice.
We therefore propose that this condition can be used to determine bulk phase coexistence in the thermodynamic limit at a NESS.

\begin{figure}[t]
    \centering\vskip-2ex
    \includegraphics[scale=0.8]{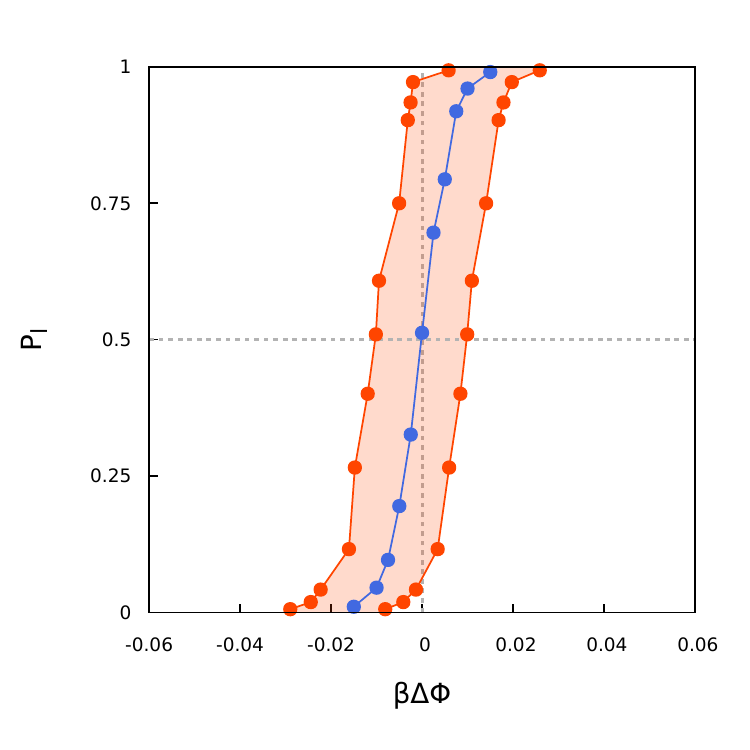}\vskip-3.5ex
    \caption{ Stochastic direct coexistence simulations verify that $\Delta\Phi = 0$ guarantees mechanical balance between coexisting phases.
      Both the nonequilibrium homogeneous (blue) and inhomogeneous (orange) cases are simulated at $\beta\Delta\mu=1.87$, where NEUS indicates that $\beta\Delta\Phi = 0$ in the thermodynamic limit.
      The right orange curve is plotted versus $\beta\Delta\Phi(L=16)$ obtained directly from NEUS simulations, while the left orange curve is plotted versus $\beta\Delta\Phi(L=\infty)$ obtained by extrapolating the NEUS simulation results to the thermodynamic limit [see Fig.~\hyperref[fig:NEUS]{S1}].
      We estimate that direct coexistence simulations performed in an infinitely large system would lie within the shaded region in between these curves.}
    \label{fig:slab}
\end{figure}

We verify that our definition of phase coexistence based on $\Delta\Phi = 0$ coincides with mechanical balance between the phases, and thus is a proper extension of the equilibrium concept, by performing stochastic direct coexistence simulations \cite{noya2008}.
We simulate two bulk phases in direct contact at a flat interface and allow the system to evolve at steady state until the lattice is fully occupied by either of the bulk phases.
Using a slab geometry on a $16\times64$ lattice, we initialize these simulations with half of the lattice in the liquid phase and the other half in the vapor phase.
If the bulk phases are in mechanical balance, then the interface should diffuse in either direction without any bias and thus reach either of the absorbing states with equal probability.
We therefore measure the probability for a system to reach the liquid phase from the initial condition, $P_\textit{l}$, to verify that unbiased diffusion coincides with $\Delta\Phi = 0$.
Fig.~\hyperref[fig:slab]{S2} shows that in the homogeneous case, $P_\textit{l} = \nicefrac{1}{2}$ at $\Delta\Phi = 0$ as expected.
Due to the finite-size effects described above, the results of these simulations in the inhomogeneous case depend on both the longitudinal and transverse dimensions of the lattice.
From the difference between $\beta\Delta\Phi(L=16)$ and $\beta\Delta\Phi(L=\infty)$ shown in Fig.~\hyperref[fig:NEUS]{S1(C)}, we are able to establish that the $P_\textit{l}(\beta\Delta\Phi)$ curve lies within the bounds shown in Fig.~\hyperref[fig:slab]{S2}.
These results are thus consistent with $\Delta\Phi = 0$ corresponding to $P_\textit{l} = \nicefrac{1}{2}$ in the inhomogeneous case.
We note that the region of uncertainty in $\beta\Delta\Phi$ shown in Fig.~\hyperref[fig:slab]{S2} (approximately $\pm0.01$) is much smaller than the magnitude of $\beta\Delta\Phi$ in all simulations used to test the applicability of classical nucleation theory ($\beta\Delta\Phi \geq 0.06$).
Taken together, these results demonstrate that our definition of nonequilibrium phase coexistence appropriately identifies the conditions for mechanical balance in both homogeneous and inhomogeneous nonequilibrium systems.

\section{Homogeneous nucleation simulations}

\subsection{Forward Flux Sampling (FFS)}

We utilize the forward flux sampling (FFS) rare-event simulation method \cite{allen2009forward} to calculate the nucleation rate starting from the vapor phase.
Using the largest cluster of bonding-state particles, $n$, as the reaction coordinate, we perform FFS using $M=64$ milestones from $n_0 = 6$ to $n_{63} = 1100$, with the spacing between consecutive milestones increasing monotonically from 3 to 50, as we advance the milestones.
We first determine the flux across the initial milestone, $\Phi_0$, from a steady-state trajectory in the vapor phase.
Likewise, the initial ensemble of configurations at $n_0$ is obtained by randomly selecting 1000 configurations at $n=n_0$ from a steady-state trajectory in the vapor phase.
We then calculate the probability $P(n_{\textit{i}+1} | n_\textit{i})$ that a trajectory launched from milestone $n_i$ reaches milestone $n_{i+1}$ before returning to the vapor phase.
To this end, we launch trajectories from each milestone $n_i$ until we obtain 1000 configurations at $n_{i+1}$.
We halt the simulation when the probability $P$ reaches unity.
Based on these probabilities and the initial flux measurement, the FFS expression for the nucleation rate density, $J$, is given as
\begin{equation} \label{eq:rate_FFS}
    J = \Phi_0\prod_{j=0}^M P(n_{j+1} | n_j).
\end{equation}

\subsection{Commitment probability and Zeldovich factor}

We compute the Zeldovich factor directly from FFS simulations by analyzing the commitment probability, $\phi(n)$.
The quantity $\phi(n)$ represents the probability that a system with nucleus size $n$ successfully completes the phase transformation into the stable liquid phase before returning to the metastable vapor phase.
We calculate $\phi(n_i)$ at each FFS milestone $n_i$ based on the milestone probabilities $P(n_{i+1}|n_i)$,
\begin{equation}
    \phi(n_i) = \prod_{j=i}^M P(n_{j+1} | n_j).
\end{equation}
The critical nucleus size $n^*$ is found where $\phi(n^*)=\nicefrac{1}{2}$, which is interpolated from the values of $\phi(n_i)$.
In the diffusive limit, the critical nucleus size coincides with the location of the top of the barrier on the (nonequilibrium) landscape $F(n) \equiv -\beta^{-1}\ln p(n)$, where $p(n)$ is the steady-state probability of observing a nucleus of size $n$ \cite{hummer2004}.
In this limit, $\phi(n)$ is given by
\begin{equation}\label{eq:committor_def}
    \phi(n) = \int_{n_\textit{v}}^n dn'\; e^{\beta F(n')} \bigg/ \int_{n_\textit{v}}^{n_\textit{l}} dn'\; e^{\beta F(n')},
\end{equation}
where $n = n_\textit{v}$ and $n = n_\textit{l}$ mark the boundaries of the transition region between the vapor and liquid phases.

\begin{figure}[b]
    \centering\vskip-2ex
    \includegraphics[scale=0.75]{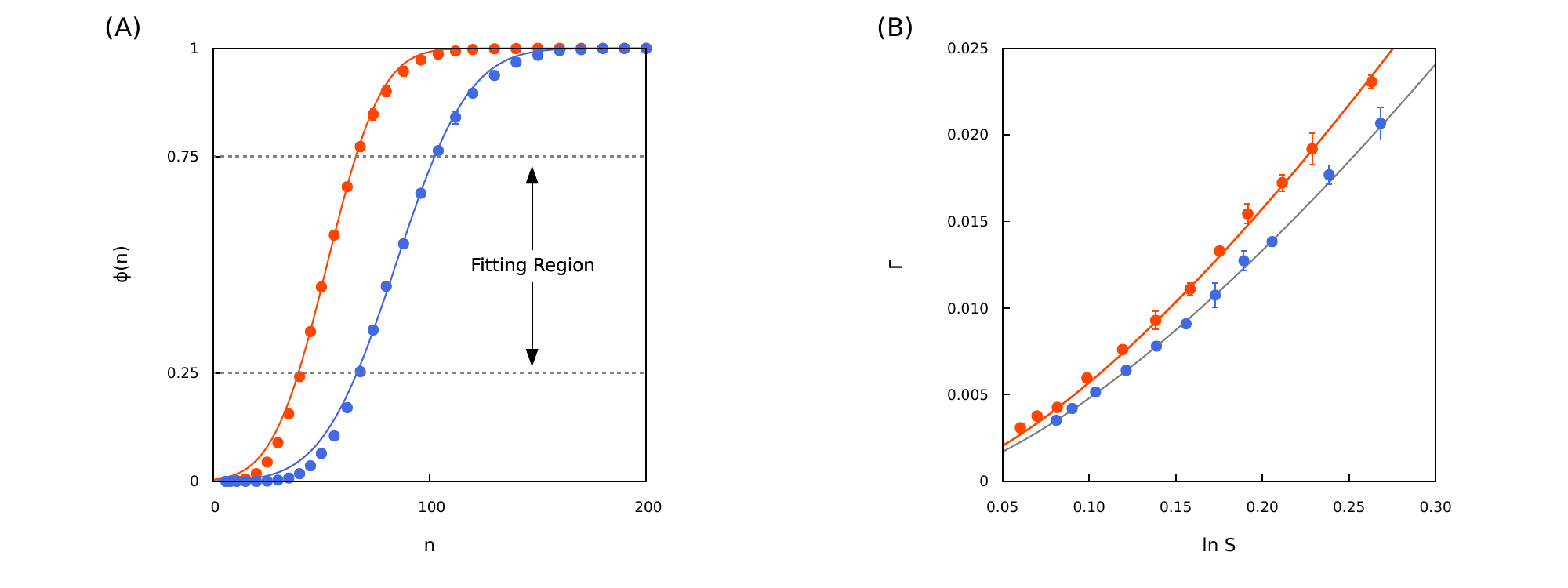}\vskip-3.5ex
    \caption{The relationship between the commitment probability, $\phi(n)$, and the Zeldovich factor, $\Gamma$. (A)~Commitment probabilities, $\phi(n)$, interpolated from FFS results (marks) and fit to \eqref{eq:commit_approx} (solid lines) at $S = 1.27$.
    (B) The Zeldovich factor, $\Gamma$, determined from the commitment probability (marks) and from the CNT kinetics equation using the analytical value $\beta\sigma_\text{eq}=1.023$ for the equilibrium line tension (gray line) and the fitted line tension $\beta\sigma = 0.856$, as described in Appendix ~\hyperref[app:noneq-line-tension]{B} (orange line). Data are shown for nonequilibrium homogeneous (blue) and inhomogeneous (orange) systems whose coexistence conditions are $\beta\Delta\mu_{\text{coex}} = 1.87$.}
    \label{fig:Zeldovich}
\end{figure}

In the high-barrier limit, the exponential integrand in \eqref{eq:committor_def} is dominated by the barrier height $\beta F(n^*)$ so that we can take saddle point approximation around $n=n^*$,
\begin{equation}
    e^{\beta F(n)} \approx e^{\beta F(n^*)}\exp\left[\dfrac{\beta F''(n^*)}{2}(n-n^*)^2\right],
\end{equation}
which leads to an approximate form of $\phi(n)$:
\begin{equation}
    \phi(n) \approx \dfrac{\erf[\sqrt{-\beta F''(n^*) / 2}(n-n^*)] + \erf[\sqrt{-\beta F''(n^*) / 2}(n^*-n_\textit{v})]}{\erf[\sqrt{-\beta F''(n^*) / 2}(n_\textit{l}-n^*)] + \erf[\sqrt{-\beta F''(n^*) / 2}(n^*-n_\textit{v})]}.
\end{equation}
The diffusive-limit condition at $\phi(n^*) = 1/2$ and the liquid-phase boundary condition $\phi(\infty) = 1$ further simplify the commitment probability into
\begin{equation}\label{eq:commit_approx}
    \phi(n) \approx \dfrac{1}{2}\erf\left[\sqrt{-\dfrac{\beta F''(n^*)}{2}}(n-n^*)\right] + \dfrac{1}{2}.
\end{equation}
Note that this approximate form of $\phi(n)$ agrees with the vapor phase boundary condition $\phi(n_\textit{v}) \approx 0$ as long as $n_\textit{v} \ll n^*$.

We evaluate the second derivative $F''(n^*)$ by fitting the commitment probabilities at the FFS milestones to \eqref{eq:commit_approx} in the region where $0.25 \leq \phi \leq 0.75$.
The Zeldovich factor, $\Gamma \equiv \sqrt{-\beta F''(n^*) / 2\pi}$, is then calculated from the fitted value of $F''(n^*)$.
Fig.~\hyperref[fig:Zeldovich]{S3(A)} shows that \eqref{eq:commit_approx} works well inside the fitting region.
Furthermore, Fig.~\hyperref[fig:Zeldovich]{S3(B)} shows that the fitted value of the Zeldovich factor and the value determined using the classical nucleation theory (CNT) line tension match one other, so that \eqref{eq:commit_approx} is consistent with CNT.

\subsection{Application of CNT to phase separation at a NESS}

\begin{figure}[!h]
    \centering\vskip-2ex
    \includegraphics[width=\textwidth]{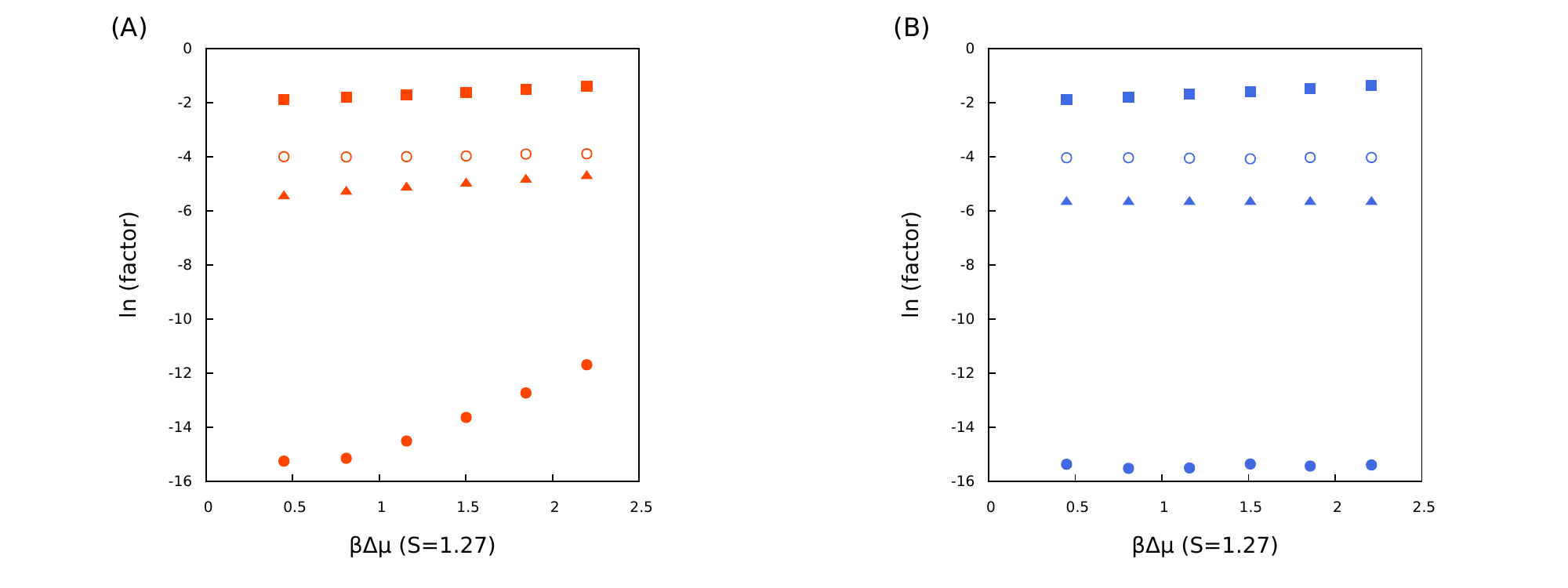}\vskip-3.5ex
    \caption{The line tension is the dominant factor governing the nucleation kinetics.
    The various factors contributing to the CNT expression for the nucleation rate in  nonequilibrium (A) inhomogeneous and (B) homogeneous cases.
    Each type of mark indicates the diffusion coefficient $D^*$ (squares), the bonding state monomer density $\rho_1$ (triangles), the Zeldovich factor $\Gamma$ (open circles), and the apparent barrier height $\ln (J/\rho_1D^*\Gamma)$ (filled circles) at $S = 1.27$.}
    \label{fig:CNT-factor-comparison}
\end{figure}

As discussed in the main text, we find that we can apply CNT to nonequilibrium systems without modifying the functional form of \eqref{eq:CNT_landscape}, although the value of the line tension may differ from the equilibrium value.
In Fig.~\hyperref[fig:CNT-factor-comparison]{S4}, we demonstrate that the line tension, $\sigma$, is the most important variable in determining the nucleation kinetics at a NESS, as expected from equilibrium systems.
Within the framework of CNT, there are three independent variables affecting the nucleation kinetics at a fixed supersaturation: $\rho_1$, $D^*$, and $\sigma$, where the last variable governs both the nucleation barrier and the Zeldovich factor.
In Fig.~\hyperref[fig:CNT-factor-comparison]{S4}, direct comparisons among $\rho_1$, $D^*$, $\Gamma$, and the apparent barrier height $\ln (J/\rho_1D^*\Gamma)$ at a fixed value of the supersaturation show that the barrier term is indeed the dominant term for both the homogeneous and inhomogeneous nonequilibrium models.
Furthermore, in contrast to the barrier term, $\Gamma$ only shows a weak dependence on $\Delta\mu$.
This demonstrates that the effect of the nonequilibrium line tension on the nucleation kinetics primarily originates from the nucleation barrier rather than the Zeldovich factor.

\section{Theoretical predictions for thermodynamics and kinetics at a NESS}
\label{sec:flex_predictions}

\subsection{Fixed Local Environment approXimation (FLEX)}
\label{sec:flex}

In FLEX, we assume that particle exchange between the open system and the reservoir relaxes to the steady state more rapidly than any change in the local configuration, or environment, around a given lattice site.
This assumption provides a good description of both phases at low temperatures arbitrarily far from equilibrium, since in this limit the number densities of particles and vacancies in the vapor and liquid phases, respectively, are extremely low.
For the two-dimensional square lattice-gas model, the local configuration comprises the four nearest-neighbor lattice sites, which determine the local potential energy $u$ when a tagged lattice site is occupied by a bonding-state particle.
Following the convention that $W_{ij}$ indicates the first-order transition rate from state $i$ to state $j$, we write the transition matrix for a single tagged lattice site with a fixed local environment as
\begin{equation}\label{eq:transition matrix}
    D^{-1}W = \begin{bmatrix}
        -(\zB+\zI) & \zB & \zI\\
        \eu & -(e^{\beta u}+D^{-1}k_\textit{BI}) & D^{-1}k_\textit{BI}\\
        1 & D^{-1}k_\textit{IB} & -(1+D^{-1}k_\textit{IB})
    \end{bmatrix},
\end{equation}
where the lattice-site states are ordered $(\text{E}, \text{B}, \text{I})$.
Solving the master equation, $0 = d\tilde{\rho}/dt = \tilde{\rho}W$, for the steady state distribution $\tilde{\rho}$ at the tagged lattice site leads to
\begin{alignat}{4}
    &\dfrac{\pB}{\pE} &&= \dfrac{\zB+\kIB(\zB+\zI)}{\eu(1+\kIB)+\kBI}
    &&= z_\text{B}\left[\dfrac{1+\kIB(1+e^{\beta\Delta f_\text{res}})}{1+\kIB(1+e^{\beta\Delta f_\text{res}+\beta\Delta\mu})}\right]e^{-\beta u}
    && \equiv \zBT(u) \exp(-\beta u)\label{eq:z_B'}\\
    &\dfrac{\pI}{\pE} &&= \dfrac{\eu\zI + \kBI(\zB+\zI)}{\eu(1+\kIB)+\kBI}
    &&= \zI \left[\dfrac{1+\kIB (1+e^{\beta\Delta f_\text{res}})e^{\beta\Delta\mu}}{1+\kIB(1+e^{\beta\Delta f_\text{res}+\beta\Delta\mu})}\right]
    && \equiv \zIT(u)\label{eq:z_I'},
\end{alignat}
where $\tilde{\rho}_i$ is the steady-state probability of being in state $i$ given the specified $u$.
The probability $\tilde{\rho}_i$ may be regarded as the number density of a particle, if $i=\textit{B}$ or \textit{I}, or a vacancy, if $i=\textit{E}$, within the FLEX framework.
\eqref{eq:z_B'} and \eqref{eq:z_I'} suggest that the steady-state distribution at the tagged lattice can be described by an effective equilibrium model with particle fugacities $\zBT(u)$ and $\zIT(u)$ and partition function $\tilde{\xi}=1+\zBT+\zIT$.
This effective equilibrium model has an identical steady-state distribution (but different probability currents) as the nonequilibrium model in the fixed local environment.

\subsection{FLEX prediction of inhomogeneous chemical reactions}
\label{sec:inhomogeneous}

We first consider the effect of the internal reaction kinetics on the thermodynamics of a fluid that phase separates into dilute (vapor) and condensed (liquid) phases.
Within FLEX, \eqref{eq:FLEX_inhomogeneity} shows that under nonequilibrium conditions ($\Delta\mu \ne 0$), the effective internal free-energy difference between B and I states in the open system, $\Delta f$, may be different from the free-energy difference in the reservoir, $\Delta f_{\text{res}}$.
As long as $\kIB$ is the same in both the vapor and liquid phases, a single effective equilibrium model provides a common description of the steady-state distribution in both phases.
However, if $\kIB$ is dependent on the local potential energy, which is on average higher in the vapor phase than in the liquid phase, then the effective equilibrium descriptions must be different in the two phases.
As a result, the steady-state density distribution in a phase-separated system with inhomogeneous chemical reactions cannot be described by an effective equilibrium model that is common to both phases.

Our simulation results indicate that these insights provided by FLEX are also useful for analyzing the full lattice model.
Motivated by the FLEX analysis, we quantify the extent of inhomogeneous chemical reactions by estimating the effective free-energy difference between the two particle internal states, $\Delta f$, from simulations of each bulk phase.
In order to calculate $\Delta f$ for a system at a nonequilibrium steady state (NESS), we start from an equilibrium grand-canonical system with particle fugacities $\zB$ and $\zI$.
A lattice configuration is defined by the identities of all the lattice sites, $\{c(\bm{r})\}$, where $c\in\{\textit{E},\textit{B},\textit{I}\}$.
In a translationally symmetric system, we can assume that a tagged particle is located at the origin, $\bm{r}=0$.
Including the empty lattice site as an ``internal state,'' E, with $z_\textit{E} = 1$, we can write the equilibrium probability of the tagged particle being in internal state $i$ as\vskip-3.5ex
\begin{align}
  p_{i(\bm{r}=0)}^{\text{eq}} &= \Xi^{-1} \sum_{\{c(\bm{r})\}} \delta\left[c(\bm{r}\!=\!0) = i\right] \; \prod_{\bm{r}} z_{c(\bm{r})} \exp\left\{ - \beta \sideset{}{'}\sum_{\bm{r},\bm{r'}} u[c(\bm{r}),c(\bm{r'})] \right\} \\
  &= \left(\frac{z_i}{z_j}\right) \Xi^{-1} \sum_{\{c(\bm{r})\}} \exp\left\{-\beta \sideset{}{''}\sum \{u[i,c(\bm{r'})] - u[j,c(\bm{r'})]\}\right\} \nonumber \\
  &\qquad \times \delta\left[c(\bm{r}\!=\!0) = j\right] \; \prod_{\bm{r}}z_{c(\bm{r})} \exp\left\{ - \beta \sideset{}{'}\sum_{\bm{r},\bm{r'}} u[c(\bm{r}),c(\bm{r'})] \right\} \\
  &= \left(\frac{z_i}{z_j}\right) \left\langle \exp\left\{-\beta \sideset{}{''}\sum \{u[i,c(\bm{r'})] - u[j,c(\bm{r'})]\}\right\} \right\rangle_{j(\bm{r}=0)} p_{j(\bm{r}=0)}^{\text{eq}}, \label{eq:semigrand-bennet-SI}
\end{align}
where $\Xi$ is the grand-canonical partition function, primed summation is over nearest-neighboring pairs (counting each unique bond once), double primed summation is over the nearest-neighbor sites $\bm{r'}$ of $\bm{r}=0$, and the angle brackets indicate an ensemble average conditioned on the particle at $\bm{r}=0$ being in the indicated internal state.
\eqref{eq:semigrand-bennet-SI} can be viewed as a Bennet acceptance ratio in the semi-grand ensemble or a generalization of the Widom insertion method.
Finally, we use \eqref{eq:semigrand-bennet-SI} to define the effective internal free-energy difference $\Delta f$ at a NESS by substituting $p^{\text{eq}}$ with the NESS distribution, $p$, and averaging over the ensemble of lattice configurations at steady state,
\begin{align}
\label{eq:deltaf1}
\begin{split}
\beta\Delta f
&= -\ln \left( \frac{p_\textit{B}}{p_\textit{I}} \right) + \ln \left\langle \exp\left\{-\beta \sideset{}{''}\sum u[\textit{B},c(\bm{r'})] - u[\textit{I},c(\bm{r'})]\right\} \right\rangle_{\textit{I}(\bm{r}=0)}\\
&= -\ln \left( \frac{p_{\textit{B}}}{p_{\textit{I}}} \right) + \ln \left\langle \exp\left\{-\beta \sideset{}{''}\sum u[\textit{B},c(\bm{r'})] \right\} \right\rangle_{\textit{I}(\bm{r}=0)},
\end{split}
\end{align}
where in the last step, we have exploited the isoenergeticity of the inert state.

\begin{figure}[!h]
    \centering
    \includegraphics[scale=0.8]{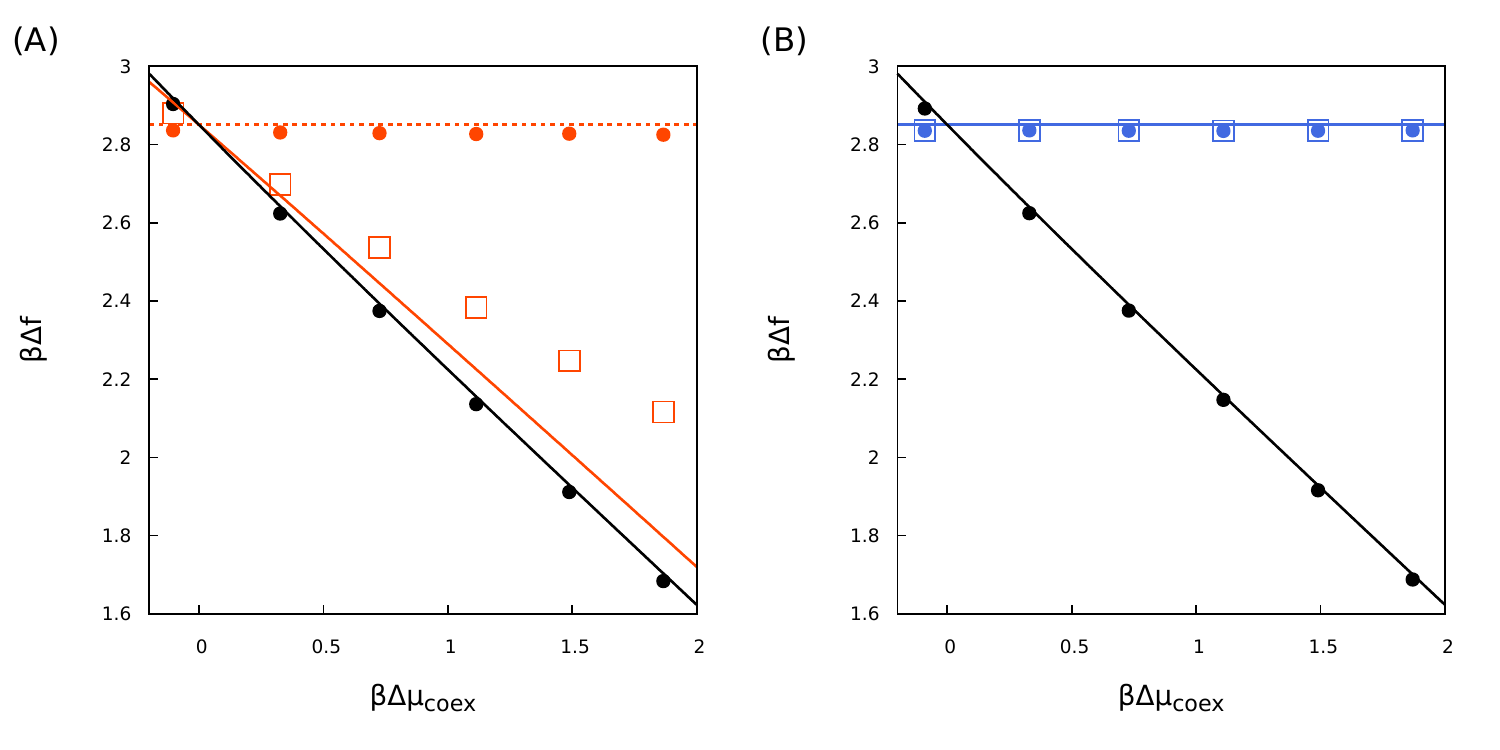}\vskip-2.5ex
    \caption{Quantification of inhomogeneous chemical reactions at a NESS.
    The effective internal free-energy differences $\Dfv$ in the vapor phase (open squares), $\Dfl$ in the liquid phase (filled colored circles), and $\Delta f_\text{res}$ in the reservoir (filled black circles) for (A) nonequilibrium inhomogeneous and (B) homogeneous models.
    Colored solid and dotted lines show the FLEX predictions for $\Dfv$ and $\Dfl$, respectively, while black lines show predictions for $\Delta f_\text{res}$ at coexistence.
    All data shown correspond to the coexistence conditions specified in Fig.~\hyperref[fig:1]{1}.}
    \label{fig:fig-Df}
\end{figure}

We numerically test the effect of applying nonequilibrium drive to the models with homogeneous and inhomogeneous chemical reactions by measuring the effective internal free-energy differences in coexisting liquid and vapor phases, $\Dfl$ and $\Dfv$, respectively.
Fig.~\hyperref[fig:fig-Df]{S5} shows the free-energy differences calculated according to \eqref{eq:deltaf1} using the NESS distribution obtained from simulations in each phase.
The measured values of the effective internal free-energy differences, $\Dfl$ and $\Dfv$, for nonequilibrium homogeneous systems are identical in the liquid and vapor phases as expected.
By contrast, the effective internal free-energy difference in the dilute phase, $\Dfv$, exhibits a monotonic decrease with respect to the nonequilibrium drive in inhomogeneous systems.
We calculate the internal free-energy differences in each phase by evaluating \eqref{eq:FLEX_inhomogeneity} at $u = 0$ and $u = 4\epsilon$ for the vapor and liquid phases, respectively.
$\Dfv$ and $\Dfl$ both show the same trend for both nonequilibrium models.
It is important to note that the existence of an effective equilibrium that is the same in both phases does not mean that a driven, homogeneous system corresponds to a true equilibrium, because the entropy production rate is always positive due to particle exchange with the reservoir.

\subsection{FLEX prediction of the entropy production rate}

\begin{figure}[!h]
    \centering
    \includegraphics[width=\textwidth]{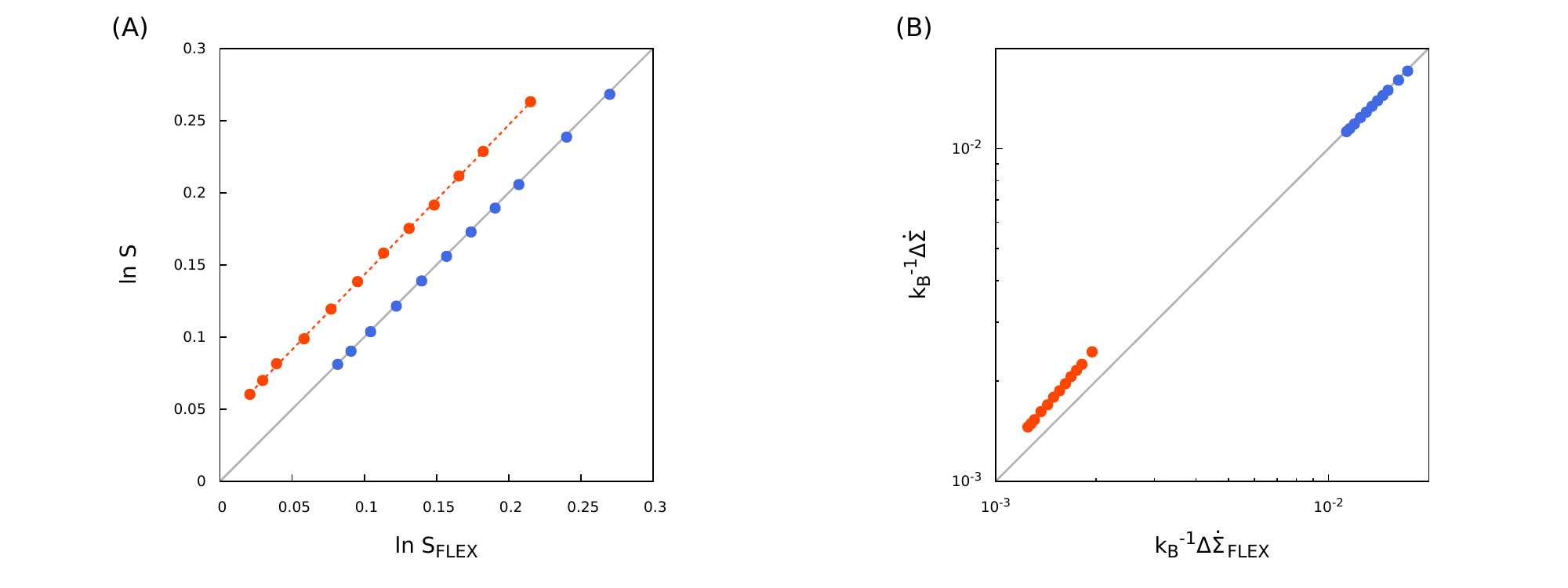}\vskip-3.5ex
    \caption{FLEX predictions of thermodynamic quantities.
    (A) Comparison between the FLEX predictions and simulation results for the supersaturation and (B) the entropy production rate density difference $\Delta\dot{\Sigma} \equiv \dot{\Sigma}_\textit{v} - \dot{\Sigma}_\textit{l}$.
    Data shown are for nonequilibrium homogeneous (blue) and inhomogeneous (orange) systems, both of whose coexistence conditions are $\beta\Delta\mu_{\text{coex}} = 1.87$.
    FLEX predictions are shown for the same conditions.}
    \label{fig:FLEX-thermodynamics}
\end{figure}

FLEX predicts the following expression for the entropy production rate density, $\dot{\Sigma}_\text{FLEX}$,
\begin{equation}
\label{eq:FLEX_entropy_production_rate}
    k_B^{-1}\dot{\Sigma}_\text{FLEX} = j\times\beta\Delta\mu
    = \dfrac{\zIT\kIB(e^{\beta\Delta\mu}-1)\beta\Delta\mu}{[1+\zBT e^{-\beta u}+\zIT][1+\kIB(1+e^{\beta\Delta f_\text{res}})e^{\beta\Delta\mu}]},
\end{equation}
where $j\equiv \pB\kBI - \pI\kIB$ is the net transition flux in the \textit{B}-to-\textit{I} direction.
\eqref{eq:FLEX_entropy_production_rate} indicates that $\dot{\Sigma}$ is always positive unless the system is at equilibrium ($\beta\Delta\mu=0$).
Furthermore, the entropy production always depends on $u$, regardless of the functional form of $\kIB$, which implies that $\dot{\Sigma}$ should in general differ between the vapor and liquid phases.
Within FLEX, we estimate the entropy production rate density in the vapor, $\dot{\Sigma}_\textit{v}$, and in the liquid phase, $\dot{\Sigma}_\textit{l}$, by fixing $u=0$ and $u=4\epsilon$, respectively. 

Fig.~\hyperref[fig:FLEX-thermodynamics]{S6} shows that the FLEX predictions for the supersaturation and entropy production rates at phase coexistence agree qualitatively with the simulation results.
The supersaturation, $S$, and the steady-state entropy production rate density, $\dot{\Sigma}$, are calculated from NEUS and the simulated trajectory as described in Ref.~\cite{van2015ensemble}, respectively.
The FLEX predictions perfectly match the simulation results for the nonequilibrium homogeneous case, while the inhomogeneous case shows systematic deviations; however, there is a clear linear relation between the predictions and the simulation results even when the system is driven far from equilibrium.

\subsection{FLEX prediction of the nonequilibrium line tension}

The emergence of a reduced effective bonding strength, $|\tilde\epsilon| \le |\epsilon|$, in our inhomogeneous simulations when $\epsilon > -\Delta f_{\text{res}} - \Delta\mu_{\text{coex}}$ can be understood using FLEX (see Appendix ~\hyperref[app:FLEX-interface]{D} in the main text).
In the bulk liquid phase, $u\approx4\epsilon$, and so $\kIB = k^\circ$ in the range of parameters that we simulated.
However, under conditions where $\epsilon > -\Delta f_\text{res} - \Delta\mu_\text{coex}$, $\kIB = k^\circ\exp({-\beta\epsilon -\beta\Delta f_\text{res} - \beta\Delta\mu_\text{coex}}) < k^\circ$ at a flat interface where a single adatom experiences a local environment of $u=\epsilon$.
This difference between the reaction kinetics in the bulk liquid and at the interface decreases $\Delta f$ at the interface, corresponding to an increase in the bonding-state particle population, $\pB$, at the interface relative to what would be expected from the effective equilibrium model for the bulk liquid at phase coexistence.
As a result, the effective adatom bonding strength, $|\tilde{\epsilon}|$, is reduced relative to $|\epsilon|$ according to \eqref{eq:effective_bond_strength}.
We note that this effect depends on both the sign of the chemical drive, $\Delta\mu$, and whether $\kIB$ is an increasing or decreasing function of $u$.

\begin{figure}[!h]
    \centering
    \includegraphics[width=0.7\textwidth]{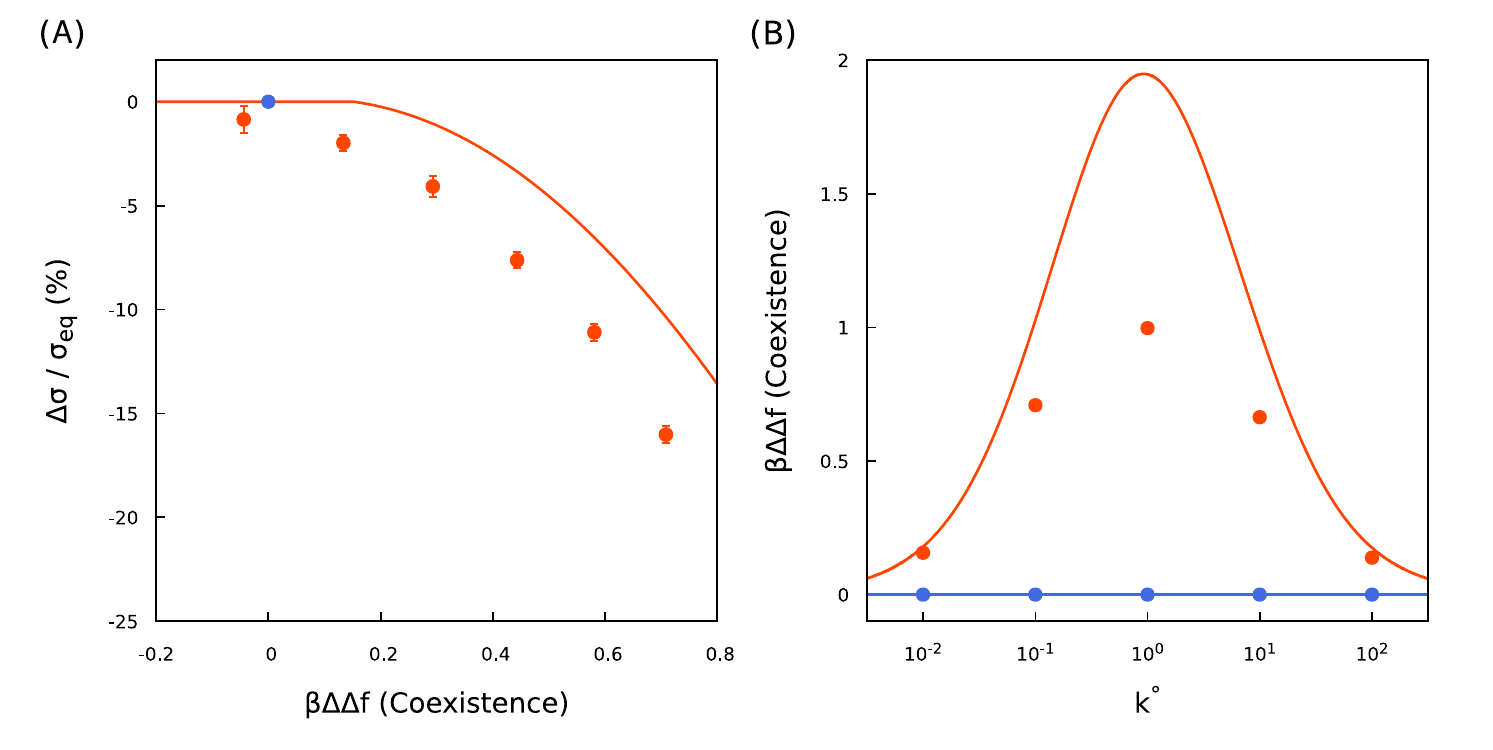}\vskip-2ex
    \caption{FLEX prediction of the nonequilibrium line tension.
    (A) The relationship between the thermodynamic inhomogeneity, $\Delta\Delta f$, at coexistence and the deviation of the line tension from equilibrium, $\Delta\sigma$.
    (B)~The dependence of $\Delta\Delta f$ on the relative timescale, $k^\circ$, at the same conditions shown in the inset of Fig.~3a in the main text.
    Solid lines show the FLEX predictions, and marks report simulation results.
    Data are shown for nonequilibrium homogeneous (blue) and inhomogeneous (orange) models.}
    \label{fig:line-tension-and-inhomogeneity}
\end{figure}

Based on this analysis, we postulate that if the inferred nonequilibrium line tension differs from the equilibrium value, then the coexisting phases at a NESS must be thermodynamically inhomogeneous and thus described by different effective equilibrium models.
Both FLEX and our simulation results support this postulated relationship between the thermodynamic inhomogeneity and the interfacial properties.
In Fig.~\hyperref[fig:line-tension-and-inhomogeneity]{S7}, the degree of inhomogeneity, $\Delta\Delta f \equiv \Dfl - \Dfv$, is calculated from simulation trajectories obtained in each phase at steady state using \eqref{eq:deltaf1}, while the FLEX predictions are calculating using the FLEX expression for $\Delta f$ and assuming that $u=4\epsilon$ for $\Dfl$ and $u=0$ for $\Dfv$.
Both simulation and theory are consistent with our prediction that a deviation in the line tension ($\Delta\sigma \neq 0$) implies a nonzero $\Delta\Delta f$.
At the same time, simulation and theory both show that the converse does not necessarily hold, as nonzero values of $\Delta\Delta f$ may not result in nonzero values of $\Delta\sigma$.
FLEX suggests that this latter relationship is dependent on the precise functional form of $\kIB$.

For the inhomogeneous model, we find that the maximum of $\beta\Delta\Delta f$ occurs when $k^\circ\approx 1$, which is also when the line tension deviates furthest from the equilibrium value (Fig.~\hyperref[fig:line-tension-and-inhomogeneity]{S7(B)}; see also the inset of \figref{3}{a}).
This observation further supports our hypothesis that changes in the interfacial properties are only possible when the two coexisting nonequilibrium phases do not share a common effective equilibrium description.
We note that, however, that the two limits $k^\circ\rightarrow0$ and $k^\circ\rightarrow\infty$ do not correspond to the same steady-state distribution.
In the limit $k^\circ\rightarrow0$, $\kIB$ also approaches zero regardless of its functional form, and the system reverts back to a true equilibrium so that $\zBT=\zB$ and $\zIT=\zI$ (\eqref{eq:z_B'} and \eqref{eq:z_I'}).
In the limit $\kIB\rightarrow\infty$, however, the fugacities in the system and the reservoir are not identical unless the system is at equilibrium ($\beta\Delta\mu = 0$).

\subsection{FLEX prediction of the nonequilibrium nucleation kinetics}

We can also derive approximate expressions for the various factors governing the nucleation kinetics, $\rho_1$, $D^*$, $\Gamma$, and $\beta\Delta F^*$, within the FLEX framework.
For the monomer density in the vapor phase, $\rho_1$, we assume that the bonding-state particles are sparsely distributed and thus the local potential energy $u$ is zero.
Then $\rho_1$ and the total particle density, $\rho_\textit{v}$, are approximated as
\begin{align}
    \rho_1 &= (\pB)_{u=0} = \dfrac{\zBT(0)}{\zBT(0)+\zIT(0)+1} \label{eq:FLEX_monomer_density}\\
    \rho_\text{v} &= (\pB+\pI)_{u=0} = \dfrac{\zBT(0)+\zIT(0)}{\zBT(0)+\zIT(0)+1}
    \label{eq:FLEX_total_particle_density},
\end{align}
where the steady-state distributions $\pB$ and $\pI$ are given by \eqref{eq:z_B'} and \eqref{eq:z_I'}.

We approximate the diffusion coefficient at the top of the nucleation barrier, $D^*$, as the rate of attaching a bonding-state adatom to a circular nucleus of size $n^*$.
We assume that the adatom interacts only with the critical nucleus and that the local environment can therefore be described by $u = \epsilon$.
Under this condition, the mean time to insert a bonding-state adatom into an unoccupied lattice site, $\tilde{T}$, is given by
\begin{equation}
  \tilde{T} = \left.\dfrac{(1+\zIT)+\kIB}{\zBT+\kIB(\zBT+\zIT)}\right|_{u=\epsilon}.
\end{equation}
We approximate the attachment as a first-order transition and define the bonding-state adatom attachment rate, $w_+$, to be the inverse of the mean time,
\begin{equation}
    w_+ \equiv \tilde{T}^{-1} =
    \left[\zBT + \zIT\times\dfrac{\kIB-\zBT}{\kIB+(1+\zIT)}\right]_{u=\epsilon}.
\end{equation}
We then approximate the diffusion coefficient as the product of the perimeter of a circular critical nucleus, $\sqrt{4\pi n^*}$, and the bonding-state adatom attachment rate per lattice site, $w_+$,
\begin{equation}
    D^* = \sqrt{4\pi n^*}w_+.
\end{equation}
Finally, given the nonequilibrium line tension $\sigma$ estimated from the approximation described above, the barrier height $\Delta F^* = F(n^*) - F(1)$ is calculated from the equilibrium barrier-height expression given in Appendix B of the main text, and the Zeldovich factor $\Gamma$ is given by
\begin{equation}
  \Gamma = \sqrt{-\dfrac{\beta F''(n^*)}{2\pi}}
  = \sqrt{\dfrac{1}{8\pi}\left[\dfrac{\beta\sigma\sqrt{4\pi}}{(n^*)^{1.5}}+\dfrac{5}{(n^*)^2}\right]}.
\end{equation}

\begin{figure}[!h]
    \centering
    \includegraphics[scale=0.8]{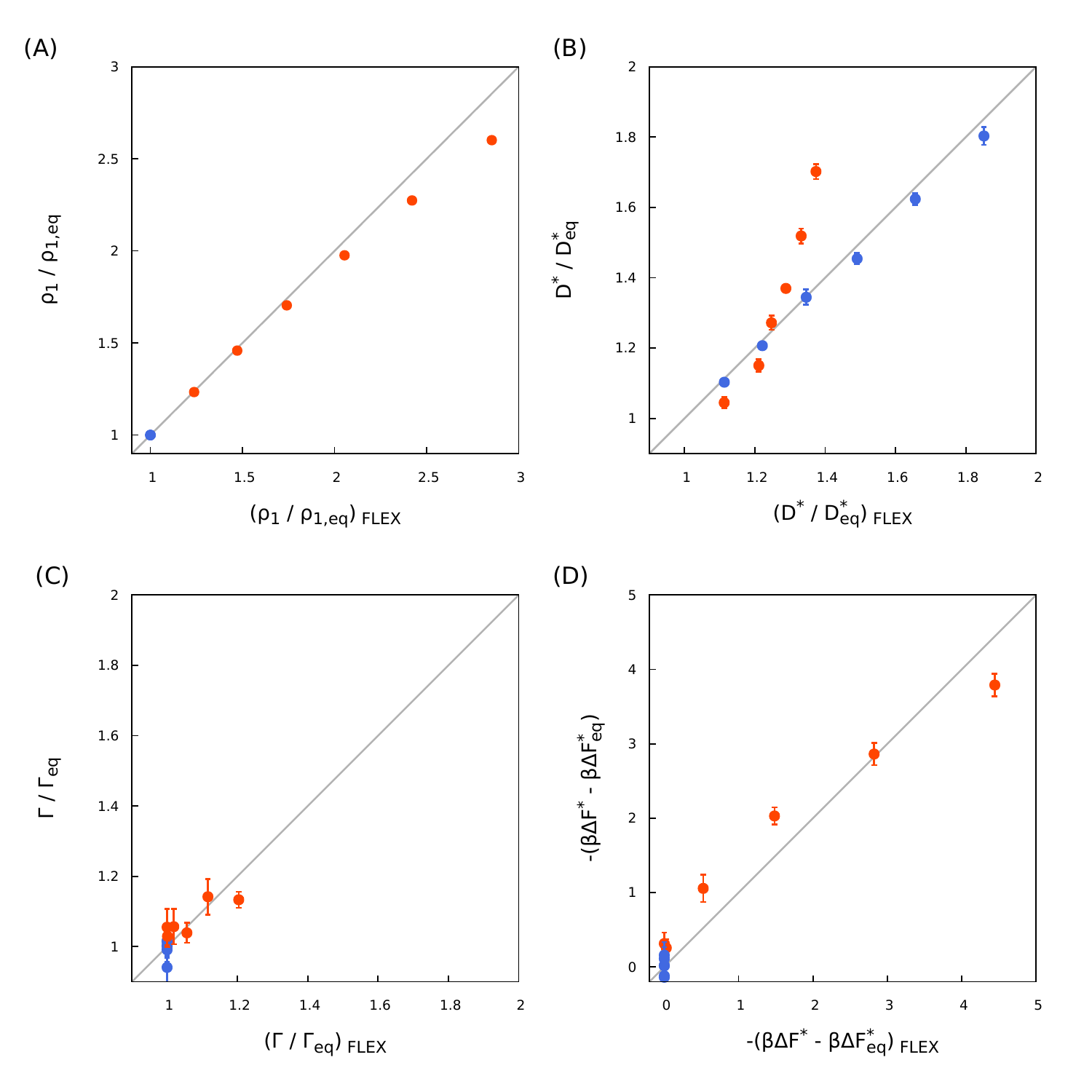}\vskip-3.5ex
    \caption{FLEX predictions for the factors governing the nucleation kinetics at a NESS.
    Comparison of the ratio between nonequilibrium and equilibrium values for
    (A) the monomer density $\rho_1$,
    (B) the diffusion coefficient $D^*$,
    (C) the Zeldovich factor $\Gamma$, and
    (D) the apparent nucleation barrier height $\beta\Delta F^*$
    at $S = 1.27$.
    FLEX predictions are made at the same $\beta\Delta\mu$ and $\rho_\textit{v}$ at $S=1.27$ for each mark displayed.
    Nonequilibrium homogeneous (blue) and inhomogeneous (orange) systems share the common coexistence condition $\beta\Delta\mu_{\text{coex}} = 1.87$.}
    \label{fig:FLEX-factor-comp-to-eq}
\end{figure}

Comparisons between these FLEX predictions and simulation results for both equilibrium and nonequilibrium systems are shown in Fig.~\hyperref[fig:FLEX-factor-comp-to-eq]{S8}.
Overall, we find qualitative agreement for all four factors in the CNT rate equation.
Importantly, FLEX qualitatively predicts the enhanced kinetics for nonequilibrium systems shown in \figref{3}{b}: Homogeneous systems show substantial enhancement only for the diffusion coefficient, while the most substantial contribution to the enhanced kinetics in the inhomogeneous case result from the apparent nucleation barrier, as indicated in Fig.~\hyperref[fig:CNT-factor-comparison]{S4(A)}.

\end{document}